\documentclass[
    ,final            
    ,numberedheadings 
  ]
  {aipproc}

\layoutstyle{6x9}

\usepackage{graphicx}
\usepackage{amsmath}
\usepackage{color}

\definecolor{darkgreen}{rgb}{0,0.5,0}
\definecolor{purple}{rgb}{0.5,0,0.5}
\definecolor{nblue}{rgb}{0.0,0.0,0.50}
\definecolor{scarlet}{rgb}{1.0,0.2,0}

\newcommand{\lsim}{\mathrel{\rlap{\lower4pt\hbox{\hskip0pt$\sim$}}
\raise1pt\hbox{$<$}}}           
\newcommand{\gsim}{\mathrel{\rlap{\lower4pt\hbox{\hskip0pt$\sim$}}
\raise1pt\hbox{$>$}}}           

\begin{document}

\title{Hadron Physics: The Essence of Matter}


\author{Lei Chang}{
  address={Institute of Applied Physics and Computational Mathematics, Beijing 100094, China}
  }

\author{Craig D.\ Roberts}{
  address={Physics Division, Argonne National Laboratory, Argonne, Illinois 60439, USA},
  altaddress={Department of Physics, Peking University, Beijing 100871, China}
  }

\begin{abstract}
Dynamical chiral symmetry breaking (DCSB) is a remarkably effective mass generating mechanism.  It is also, amongst other things, the foundation for a successful application of chiral effective field theories, the origin of constituent-quark masses, and intimately connected with confinement in QCD.  Using the Dyson-Schwinger equations (DSEs), we explain the origin and nature of DCSB, and elucidate some of its consequences, e.g.: a model-independent result for the pion susceptibility; the generation of a quark anomalous chromomagnetic moment, which may explain the longstanding puzzle of the $a_1$-$\rho$ mass splitting; its impact on the behaviour of the electromagnetic pion form factor -- thereby illustrating how data can be used to chart the momentum-dependence of the dressed-quark mass function; in the form of the pion and kaon valence-quark parton distribution functions, and the relation between them; and aspects of the neutron's electromagnetic form factors, in particular $F_1^u/F_1^d$ and $G_M^n$.  We argue that in solving QCD, a constructive feedback between theory and extant and forthcoming experiments will most rapidly enable constraints to be placed on the infrared behaviour of QCD's $\beta$-function, the nonperturbative quantity at the core of hadron physics; and emphasise throughout the role played by confrontation with data as a means of verifying our understanding of Nature.
\end{abstract}

\maketitle


\section{Introduction}
A hundred years and more of fundamental research in atomic and nuclear physics has shown us that all matter is corpuscular, with the atoms that comprise us, themselves containing a dense nuclear core.  This core is composed of protons and neutrons, referred to collectively as nucleons, which are members of a broader class of femtometre-scale particles, called hadrons.  In working toward an understanding of hadrons, we have discovered that they are complicated bound-states of quarks and gluons, which are elementary and pointlike excitations whose interactions are described by a Poincar\'e invariant quantum non-Abelian gauge field theory; namely, quantum chromodynamics (QCD).  The goal of hadron physics is the provision of a quantitative explanation of the properties of hadrons through a solution of QCD.

Quantum chromodynamics is the strong-interaction part of the Standard Model of Particle Physics and solving QCD presents a fundamental problem that is unique in the history of science.  Never before have we been confronted by a theory whose elementary excitations are not those degrees-of-freedom readily accessible via experiment; i.e., whose elementary excitations are confined.  Moreover, we have numerous reasons to believe that QCD generates forces which are so strong that less-than 2\% of a nucleon's mass can be attributed to the so-called current-quark masses that appear in the QCD Lagrangian; viz., forces that generate mass from almost nothing, a phenomenon known as dynamical chiral symmetry breaking (DCSB).

Neither confinement nor DCSB is apparent in QCD's Lagrangian and yet they play the dominant role in determining the observable characteristics of real-world QCD.  The physics of hadrons is ruled by \emph{emergent phenomena}, such as these, which can only be elucidated through the employment of nonperturbative methods in quantum field theory.  This is both the greatest novelty and the greatest challenge within the Standard Model.  We must find essentially new ways and means to explain precisely via mathematics the observable content of QCD.

This contribution to these Proceedings provides an introduction to hadron physics and a review of selected recent progress in this field made using QCD's Dyson-Schwinger equations (DSEs).  The complex of DSEs is a powerful tool, which has been employed with marked success to study confinement and DCSB, and their impact on hadron observables.  This is apparent from the detailed background material that is available in Refs.\,\cite{Roberts:1994dr,Roberts:2000aa,Alkofer:2000wg,lecturenotes,Maris:2003vk,%
Holl:2006ni,Fischer:2006ub,Roberts:2007jh,Roberts:2007ji,Holt:2010vj,Kronfeld:2010bx} and will be exemplified herein.

\section{Hadron Physics -- Some Key Points}
The basic problem of hadron physics is to solve QCD.  This inspiring goal will only be achieved through a joint effort from experiment and theory.  The hadron physics community now has a range of major facilities that are accumulating data, of unprecedented accuracy and precision, which pose important challenges for theory.  Notable amongst these facilities is the Thomas Jefferson National Accelerator Facility (JLab) in Newport News, Virginia.  The interpretation and prediction of phenomena at JLab provide the themes for much of the material presented herein.  In hadron physics, it is the feedback between experiment and theory that leads most rapidly to progress in understanding.  The opportunities for researchers in hadron physics promise to grow because upgraded and new facilities will appear on a five-to-ten-year time-scale \cite{12GeVJLab,JPARC,kpeters,eicc}.

Asymptotic coloured states have not been observed, but is it a cardinal fact that they cannot?  No solution to QCD will be complete if it does not explain confinement.  This means confinement in the real world, which contains quarks with light current-quark masses.  This is distinct from the artificial universe of pure-gauge QCD without dynamical quarks, studies of which tend merely to focus on achieving an area law for a Wilson loop and hence are irrelevant to the question of light-quark confinement.

In stepping toward an answer to the question of confinement, it will likely be necessary to map out the long-range behaviour of the interaction between light-quarks; namely, QCD's $\beta$-function at infrared momenta.  In this connection we note that the spectrum of meson and baryon excited states, and hadron elastic and transition form factors provide unique information about the long-range interaction between light-quarks and, in addition, the distribution of a hadron's characterising properties -- such as mass and momentum, linear and angular -- amongst its QCD constituents.  The upgraded and promised future facilities will provide data that should guide the charting process.   However, to make full use of that data, it will be necessary to have Poincar\'e covariant theoretical tools that enable the reliable study of hadrons in the mass range $1$-$2\,$GeV.  Crucially, on this domain both confinement and DCSB are germane.

\begin{figure}[t]

\centerline{
\includegraphics[clip,width=0.4\textheight]{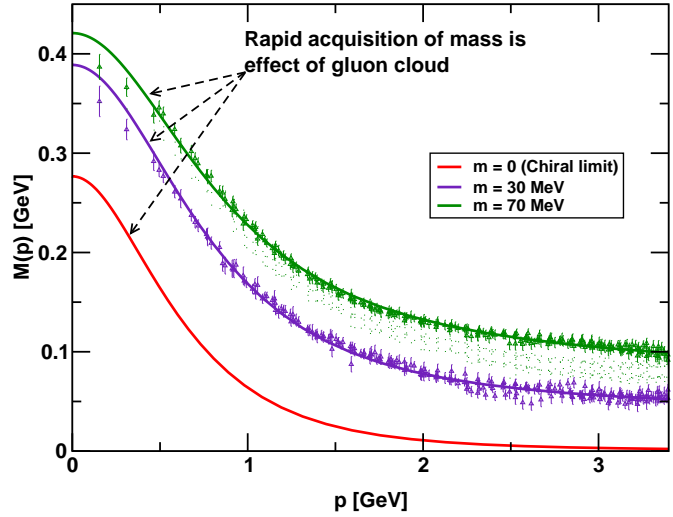}}

\caption{\label{gluoncloud}
Dressed-quark mass function, $M(p)$ in Eq.\,(\protect\ref{Sgeneral}): \emph{solid curves} -- DSE results, obtained as explained in \protect\cite{Bhagwat:2003vw,Bhagwat:2006tu}, ``data'' -- numerical simulations of lattice-QCD \protect\cite{Bowman:2005vx}.  (NB.\ $m=70\,$MeV is the uppermost curve and current-quark mass decreases from top to bottom.)  One observes the current-quark of perturbative QCD evolving into a constituent-quark as its momentum becomes smaller.  The constituent-quark mass arises from a cloud of low-momentum gluons attaching themselves to the current-quark.  This is dynamical chiral symmetry breaking (DCSB): an essentially nonperturbative effect that generates a quark mass \emph{from nothing}; namely, it occurs even in the chiral limit.  
(Figure adapted from Ref.\,\protect\cite{Bhagwat:2007vx}.)}
\end{figure}

It is known that DCSB; namely, the generation of mass \emph{from nothing}, does take place in QCD.  It arises primarily because a dense cloud of gluons comes to clothe a low-momentum quark \cite{Bhagwat:2007vx}.  This is best seen by solving the DSE for the dressed-quark propagator; i.e., the gap equation, which yields the result illustrated in Fig.\,\ref{gluoncloud}.  However, the origin of the interaction strength at infrared momenta, which guarantees DCSB through the gap equation, is currently unknown.  This relationship ties confinement to DCSB.  The reality of DCSB means that the Higgs mechanism is largely irrelevant to the bulk of normal matter in the universe.  Instead the single most important mass generating mechanism for light-quark hadrons is the strong interaction effect of DCSB; e.g., one can identify it as being responsible for 98\% of a proton's mass.

It is natural to ask whether the connection between confinement and DCSB is accidental or causal.  There are models with DCSB but not confinement, however, a model with confinement but lacking DCSB has not yet been identified (see, e.g., Secs.\,2.1 and 2.2 of Ref.\,\cite{Roberts:2007jh}).  This leads to a conjecture that DCSB is a necessary consequence of confinement.  It is interesting that there are numerous models and theories which exhibit both confinement and DCSB, and possess an external control parameter such that deconfinement and chiral symmetry restoration occur simultaneously at some critical value of this parameter; e.g., quantum electrodynamics in three dimensions with $N_f$ electrons \cite{Bashir:2008fk,Bashir:2009fv}, and models of QCD at nonzero temperature and chemical potential \cite{Bender:1996bm,Fischer:2009gk,Blaschke:1997bj,Bender:1997jf,Chen:2008zr}.  Whether this simultaneity is a property possessed by QCD, and/or some broader class of theories, in response to changes in: the number of light-quark flavours; temperature; or chemical potential, is a longstanding question.

The momentum-dependence of the quark mass, illustrated in Fig.\,\ref{gluoncloud}, is an essentially quantum field theoretic effect, unrealisable in quantum mechanics, and a fundamental feature of QCD.  This single curve connects the infrared and ultraviolet regimes of the theory, and establishes that the constituent-quark and current-quark masses are simply two connected points separated by a large momentum interval.  The curve shows that QCD's dressed-quark behaves as a constituent-quark, a current-quark, or something in between, depending on the momentum of the probe which explores the bound-state containing the dressed-quark.  It follows that calculations addressing momentum transfers $Q^2 \gsim M^2$, where $M$ is the mass of the hadron involved, require a Poincar\'e-covariant approach that can veraciously realise quantum field theoretical effects \cite{Cloet:2008re}.  Owing to the vector-exchange character of QCD, covariance also guarantees the existence of nonzero quark orbital angular momentum in a hadron's rest-frame \cite{Roberts:2007ji,Bhagwat:2006xi,Bhagwat:2006pu}.

The dressed-quark mass function has a remarkable capacity to correlate and be party to explaining a wide range of diverse phenomena.  This brings urgency to the need to understand the relationship between parton properties in the light-front frame, whose peculiar properties simplify some theoretical analyses, and the structure of hadrons as measured in the rest frame or other smoothly related frames.  This is a problem because, e.g., DCSB, an established keystone of low-energy QCD, has not been realised in the light-front formulation.  The obstacle is the constraint $k^+:=k^0+k^3>0$ for massive quanta on the light front \cite{Brodsky:1991ir}.  It is therefore impossible to make zero momentum Fock states that contain particles and hence the vacuum is trivial.  Only the zero modes of light-front quantisation can dress the ground state but little progress has been made with understanding just how that might occur.
On the other hand, it is conceivable that this dressing is inextricably tied with the formation and structure of Goldstone modes and not otherwise a measurable property of the vacuum.  This conjecture is being explored \cite{Brodsky:2009zd,BRST10}.
In addition, parton distribution functions, which have a probability interpretation in the infinite momentum frame, must be calculated in order to comprehend their content: parametrisation is insufficient.  It would be very interesting to know, e.g., how, if at all, the distribution functions of a Goldstone mode differ from those of other hadrons \cite{Holt:2010vj}.

\section{Confinement}
\label{Sect:Conf}
It is worth stating plainly that the potential between infinitely-heavy quarks measured in numerical simulations of quenched lattice-regularised QCD -- the so-called static potential -- is simply \emph{irrelevant} to the question of confinement in the real world, in which light quarks are ubiquitous.  In fact, it is a basic feature of QCD that light-particle creation and annihilation effects are essentially nonperturbative and therefore it is impossible in principle to compute a potential between two light quarks \cite{Bali:2005fu,Chang:2009ae}.

A perspective on confinement was laid out in Ref.\,\cite{Krein:1990sf}.  Confinement can be related to the analytic properties of QCD's Schwinger functions, which are often called Euclidean-space Green functions.  For example, it can be read from the reconstruction theorem \cite{SW80,GJ81} that the only Schwinger functions which can be associated with expectation values in the Hilbert space of observables; namely, the set of measurable expectation values, are those that satisfy the axiom of reflection positivity.  This is an extremely tight constraint.  It can be shown to require as a necessary condition that the Fourier transform of the momentum-space Schwinger function is a positive-definite function of its arguments, and is discussed and illustrated in Sec.~2 of Ref.\,\cite{Roberts:2007ji}.

From this standpoint the question of light-quark confinement can be translated into the challenge of charting the infrared behavior of QCD's \emph{universal} $\beta$-function.  (Although this function may depend on the scheme chosen to renormalise the theory, it is unique within a given scheme \protect\cite{Celmaster:1979km}.  Of course, the behaviour of the $\beta$-function on the perturbative domain is well known.)  This is a well-posed problem whose solution is an elemental goal of modern hadron physics and which can be addressed in any framework enabling the nonperturbative evaluation of renormalisation constants.

\section{Nonperturbative DSE truncations}
\label{spectrum1}
\subsection{Existence and exact results}
Through the gap and Bethe-Salpeter equations (BSEs) the pointwise behaviour of the $\beta$-function determines the pattern of chiral symmetry breaking; e.g., the behaviour in Fig.\,\ref{gluoncloud}.  Moreover, the fact that these and other DSEs connect the $\beta$-function to experimental observables entails, e.g., that comparison between computations and observations of the hadron mass spectrum, and hadron elastic and transition form factors, can be used to chart the $\beta$-function's long-range behaviour.

In order to realise this goal, a nonperturbative symmetry-preserving DSE truncation is necessary.  Steady quantitative progress can be made with a scheme that is systematically improvable \cite{Munczek:1994zz,Bender:1996bb}.  In fact, the mere existence of such a scheme has enabled the proof of exact nonperturbative results in QCD.  For example, there are veracious statements about radially-excited and hybrid pseudoscalar mesons \cite{Holl:2004fr,Holl:2005vu}; and regarding the $\eta$-$\eta^\prime$ complex and $\pi^0$-$\eta$-$\eta^\prime$ mixing, with predictions of $\theta_{\eta \eta^\prime} = -15^\circ$ and $\theta_{\pi^0 \eta} = 1^\circ$ \cite{Bhagwat:2007ha}.  In connection with these systems, only theoretical studies that are demonstrably consistent with the results proved in Refs.\,\cite{Holl:2004fr,Holl:2005vu,Bhagwat:2007ha} can be considered seriously.

In addition, a novel result for the pion susceptibility was recently obtained via the isovector-pseudoscalar vacuum polarisation \cite{Chang:2009at}; viz., in the neighbourhood of the chiral-limit the pion susceptibility can be expressed as a sum of two independent terms:
\begin{equation}
\label{main}
{\cal X}_5(\zeta) \stackrel{\hat m ~ 0}{=}  - \, \frac{\langle \bar q q\rangle^0_\zeta}{m(\zeta)}  + {\cal X}(\zeta) + \mbox{O}(\hat m)\,.
\end{equation}
The first expresses the pion-pole contribution and involves the so-called vacuum-quark condensate.  The second is identical to the vacuum chiral susceptibility, which describes the response of QCD's ground-state to a fluctuation in the current-quark mass.  

Equation\,(\ref{main}) is a remarkable result, which is nonetheless readily understood.  Recall that in the absence of a current-quark mass, the two-flavour QCD action has a SU$_L(2)\otimes\,$SU$_R(2)$ symmetry; and, moreover, that ascribing scalar-isoscalar quantum numbers to the QCD vacuum is a convention, contingent upon the form of the current-quark mass term.  It follows that the massless action cannot distinguish between the continuum of sources specified by
\begin{equation}
{\rm constant} \times \int d^4 x\;  \bar q(x) \, {\rm e}^{i \gamma_5 \vec{\tau}\cdot \vec{\theta}} q(x)\,,\; |\theta|\in[0,2\pi)\,.
\end{equation}
Hence, the regular part of the vacuum susceptibility must be identical when measured as the response to any one of these sources, so that ${\cal X}_{\rm reg.}={\cal X}$ for all choices of $\vec{\theta}$.  This is the content of the so-called ``Mexican hat'' potential, which is used in building models for QCD.
The magnitude of ${\cal X}$ depends on whether the chiral symmetry is dynamically broken, or not; and the strength of the interaction as measured with respect to the critical value required for DCSB \cite{Chang:2008ec}.  When the symmetry is dynamically broken, then the Goldstone modes appear, by convention, in the pseudoscalar-isovector channel, and thus the pole contributions appear in ${\cal X}_5$ but not in the chiral susceptibility.  It is valid to draw an analogy with the Weinberg sum rule \cite{Weinberg:1967kj,Chang:2008sp}.

\subsection{Expressing DCSB in the Bethe-Salpeter kernel}
Despite the successes achieved with the systematic scheme, one anticipates that significant qualitative advances in understanding the essence of QCD could be made with symmetry-preserving kernel \emph{Ans\"atze} that expresses important additional nonperturbative effects, which are impossible to capture in any finite sum of contributions.  Such an approach has recently become available \cite{Chang:2009zb} and is worth summarising herein.

Consider, e.g., pseudoscalar and axial-vector mesons, which appear as poles in the inhomogeneous BSE for the axial-vector vertex, $\Gamma_{5\mu}^{fg}$, where $f,g$ are flavour labels.  An exact form of that equation is ($k$, $q$ are relative momenta, $P$ is the total momentum flowing into the vertex, and $q_\pm = q\pm P/2$, etc.)
\begin{eqnarray}
\nonumber
\Gamma_{5\mu}^{fg}(k;P) & = & Z_2 \gamma_5\gamma_\mu
- \int_q^\Lambda g^2 D_{\alpha\beta}(k-q)\,\frac{\lambda^a}{2}\,\gamma_\alpha S_f(q_+) \Gamma_{5\mu}^{fg}(q;P) S_g(q_-) \frac{\lambda^a}{2}\,\Gamma_\beta^g(q_-,k_-) \\
&&  + \int^\Lambda_q g^2D_{\alpha\beta}(k-q)\, \frac{\lambda^a}{2}\,\gamma_\alpha S_f(q_+) \frac{\lambda^a}{2} \Lambda_{5\mu\beta}^{fg}(k,q;P), \label{genbse}
\end{eqnarray}
where $\Lambda_{5\mu\beta}^{fg}$ is a 4-point Schwinger function that is completely defined via the inverse of the dressed-quark propagator; namely, the kernel of the gap equation:\footnote{In our Euclidean metric:  $\{\gamma_\mu,\gamma_\nu\} = 2\delta_{\mu\nu}$; $\gamma_\mu^\dagger = \gamma_\mu$; $\gamma_5= \gamma_4\gamma_1\gamma_2\gamma_3$; $\sigma_{\mu\nu}=(i/2)[\gamma_\mu,\gamma_\nu]$; $a \cdot b = \sum_{i=1}^4 a_i b_i$; and $P_\mu$ timelike $\Rightarrow$ $P^2<0$.  More information is available, e.g., in App.\,A of Ref.\,\protect\cite{Holl:2006ni}.}
\begin{eqnarray}
S(p)^{-1} =  Z_2 \,(i\gamma\cdot p + m^{\rm bm})+   Z_1 \int^\Lambda_q\! g^2 D_{\mu\nu}(p-q) \frac{\lambda^a}{2}\gamma_\mu S(q) \Gamma^a_\nu(q,p) , \label{gendse}
\end{eqnarray}
where $\int^\Lambda_q$ indicates a Poincar\'e invariant regularisation of the integral, with $\Lambda$ the regularisation mass-scale, $D_{\mu\nu}$ is the renormalised dressed-gluon propagator, $\Gamma_\nu^a$ is the renormalised dressed-quark-gluon vertex, and $m^{\rm bm}$ is the quark's $\Lambda$-dependent bare current-mass.  The vertex and quark wave-function renormalisation constants, $Z_{1,2}(\zeta^2,\Lambda^2)$, depend on the gauge parameter.  The solution to Eq.\,(\ref{gendse}) has the form
\begin{eqnarray}
 S(p) & =&
%
\frac{1}{i \gamma\cdot p \, A(p^2,\zeta^2) + B(p^2,\zeta^2)}
= \frac{Z(p^2,\zeta^2)}{i\gamma\cdot p + M(p^2)}\,,
%
\label{Sgeneral}
\end{eqnarray}
where the mass function, $M(p^2)=B(p^2,\zeta^2)/A(p^2,\zeta^2)$ is independent of the renormalisation point, $\zeta$.
The pseudoscalar vertex satisfies an analogue of Eq.\,(\ref{genbse}) and has the general form
\begin{equation}
i\Gamma_{5}^{fg}(k;P) = \gamma_5 \left[ i E_5(k;P) + \gamma\cdot P F_5(k;P) + \gamma\cdot k \, G_5(k;P) + \sigma_{\mu\nu} k_\mu P_\nu H_5(k;P) \right].
\label{genG5}
\end{equation}

In any dependable study of light-quark hadrons the solution of Eq.\,(\ref{genbse}) must satisfy the axial-vector Ward-Takahashi; viz.,
\begin{equation}
P_\mu \Gamma_{5\mu}^{fg}(k;P) + \, i\,[m_f(\zeta)+m_g(\zeta)] \,\Gamma_5^{fg}(k;P)
= S_f^{-1}(k_+) i \gamma_5 +  i \gamma_5 S_g^{-1}(k_-) \,,
\label{avwtim}
\end{equation}
which expresses chiral symmetry and its breaking pattern.  The condition
\begin{equation}
P_\mu \Lambda_{5\mu\beta}^{fg}(k,q;P) + i [m_f(\zeta)+m_g(\zeta)] \Lambda_{5\beta}^{fg}(k,q;P)
= \Gamma_\beta^f(q_+,k_+) \, i\gamma_5+ i\gamma_5 \, \Gamma_\beta^g(q_-,k_-) \,, \label{LavwtiGamma}
\end{equation}
where $\Lambda_{5\beta}^{fg}$ is the analogue of $\Lambda_{5\mu\beta}^{fg}$ in the pseudoscalar equation, is necessary and sufficient to ensure the Ward-Takahashi identity is satisfied \cite{Chang:2009zb}.

Consider Eq.\,(\ref{LavwtiGamma}).  Rainbow-ladder is the lead\-ing-or\-der term in a systematic DSE truncation scheme \cite{Munczek:1994zz,Bender:1996bb}.  It corresponds to $\Gamma_\nu^f=\gamma_\nu$, in which case Eq.\,(\ref{LavwtiGamma}) is solved by $\Lambda_{5\mu\beta}^{fg}\equiv 0 \equiv \Lambda_{5\beta}^{fg}$.  This is the solution that indeed provides the rainbow-ladder forms of Eq.\,(\ref{genbse}).  Such consistency will be apparent in any valid systematic term-by-term improvement of the rainbow-ladder truncation.

However, Eq.\,(\ref{LavwtiGamma}) is far more than just a device for checking a truncation's consistency.  For, just as the vector Ward-Takahashi identity has long been used to build \emph{Ans\"atze} for the dressed-quark-photon vertex \cite{Roberts:1994dr,Ball:1980ay,Kizilersu:2009kg}, Eq.\,(\ref{LavwtiGamma}) provides a tool for constructing a symmetry-preserving kernel of the BSE that is matched to any reasonable \emph{Ansatz} for the dressed-quark-gluon vertex which appears in the gap equation.  With this powerful capacity, Eq.\,(\ref{LavwtiGamma}) achieves a goal that has been sought ever since the Bethe-Salpeter equation was introduced \cite{Salpeter:1951sz}.  As we shall indicate, the symmetry-preserving kernel it can provide promises to enable the first reliable Poincar\'e invariant calculation of the spectrum of mesons with masses larger than 1\,GeV.

One can illustrate the utility of Eq.\,(\ref{LavwtiGamma}) through an application to ground state pseudoscalar and scalar mesons composed of equal-mass $u$- and $d$-quarks.  To this end, suppose that in Eq.\,(\ref{gendse}) one employs an \emph{Ansatz} for the quark-gluon vertex which satisfies
\begin{equation}
P_\mu i \Gamma_\mu^f(k_+,k_-) = {\cal B}(P^2)\left[ S_f^{-1}(k_+) - S_f^{-1}(k_-)\right]\,, \label{wtiAnsatz}
\end{equation}
with ${\cal B}$ flavour-independent.  (NB.\ While the true quark-gluon vertex does not satisfy this identity, owing to the form of the Slavnov-Taylor identity which it does satisfy, it is plausible that a solution of Eq.\,(\protect\ref{wtiAnsatz}) can provide a reasonable pointwise approximation to the true vertex.)  Given Eq.\,(\ref{wtiAnsatz}), then Eq.\,(\ref{LavwtiGamma}) entails ($l=q-k$)
\begin{equation}
i l_\beta \Lambda_{5\beta}^{fg}(k,q;P) =
{\cal B}(l)^2\left[ \Gamma_{5}^{fg}(q;P) - \Gamma_{5}^{fg}(k;P)\right], \label{L5beta}
\end{equation}
with an analogous equation for $P_\mu l_\beta i\Lambda_{5\mu\beta}^{fg}(k,q;P)$.  This identity can be solved to obtain
\begin{equation}
\Lambda_{5\beta}^{fg}(k,q;P)  :=  {\cal B}((k-q)^2)\, \gamma_5\,\overline{ \Lambda}_{\beta}^{fg}(k,q;P) \,, \label{AnsatzL5a}
\end{equation}
with, using Eq.\,(\ref{genG5}),
\begin{eqnarray}
\nonumber
\lefteqn{\overline{ \Lambda}_{\beta}^{fg}(k,q;P) =
2 \ell_\beta \, [ i \Delta_{E_5}(q,k;P)+ \gamma\cdot P \Delta_{F_5}(q,k;P) ]
+  \gamma_\beta \, \Sigma_{G_5}(q,k;P)} \\
&&+  2 \ell_\beta \,  \gamma\cdot\ell\, \Delta_{G_5}(q,k;P)  + [ \gamma_\beta,\gamma\cdot P]
\Sigma_{H_5}(q,k;P) + 2 \ell_\beta  [ \gamma\cdot\ell ,\gamma\cdot P]  \Delta_{H_5}(q,k;P) \,,
\label{AnsatzL5b}
\end{eqnarray}
where $\ell=(q+k)/2$, $\Sigma_{\Phi}(q,k;P) = [\Phi(q;P)+\Phi(k;P)]/2$ and $\Delta_{\Phi}(q,k;P) = [\Phi(q;P)-\Phi(k;P)]/[q^2-k^2]$.

Now, given any \emph{Ansatz} for the quark-gluon vertex that satisfies Eq.\,(\ref{wtiAnsatz}), then the pseudoscalar analogue of Eq.\,(\ref{genbse}) and Eqs.\,(\ref{gendse}), (\ref{AnsatzL5a}), (\ref{AnsatzL5b}) provide a symmetry-preserving closed system whose solution predicts the properties of pseudoscalar mesons.
The relevant scalar meson equations are readily derived.  (NB.\ We are aware of the role played by resonant contributions to the kernel in the scalar channel \protect\cite{Holl:2005st} but they are not pertinent to this discussion.)
With these systems one can anticipate, elucidate and understand the influence on hadron properties of the rich nonperturbative structure expected of the fully-dressed quark-gluon vertex in QCD: in particular, that of the dynamically generated dressed-quark mass function, whose impact is quashed at any finite order in the truncation scheme of Ref.\,\cite{Bender:1996bb}.

To proceed one need only specify the gap equation's kernel because, as noted above, the BSEs are completely defined therefrom.  To complete the illustration \cite{Chang:2009zb} a simplified form of the effective interaction in Ref.\,\cite{Maris:1997tm} was employed and two vertex \emph{Ans\"atze} were compared; viz., the bare vertex $\Gamma_\mu^g = \gamma_\mu$, which defines the rainbow-ladder truncation of the DSEs and omits vertex dressing; and the Ball-Chiu (BC) vertex \cite{Ball:1980ay} which nonperturbatively incorporates vertex dressing associated with DCSB:
\begin{equation}
i\Gamma^g_\mu(q,k)  =
i\Sigma_{A^g}(q^2,k^2)\,\gamma_\mu +
2 \ell_\mu \left[i\gamma\cdot \ell \,
\Delta_{A^g}(q^2,k^2) + \Delta_{B^g}(q^2,k^2)\right] \!.
\label{bcvtx}
\end{equation}

\begin{figure}[t]
\vspace*{-7ex}

\includegraphics[clip,width=0.35\textheight]{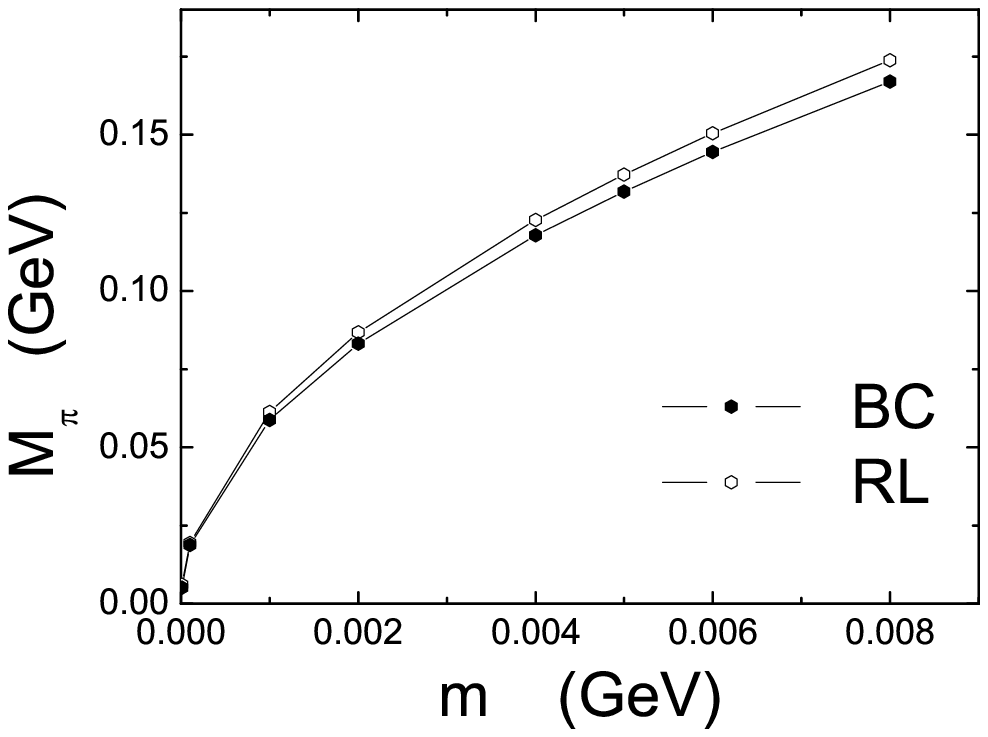}
\vspace*{-7ex}

\includegraphics[clip,width=0.35\textheight]{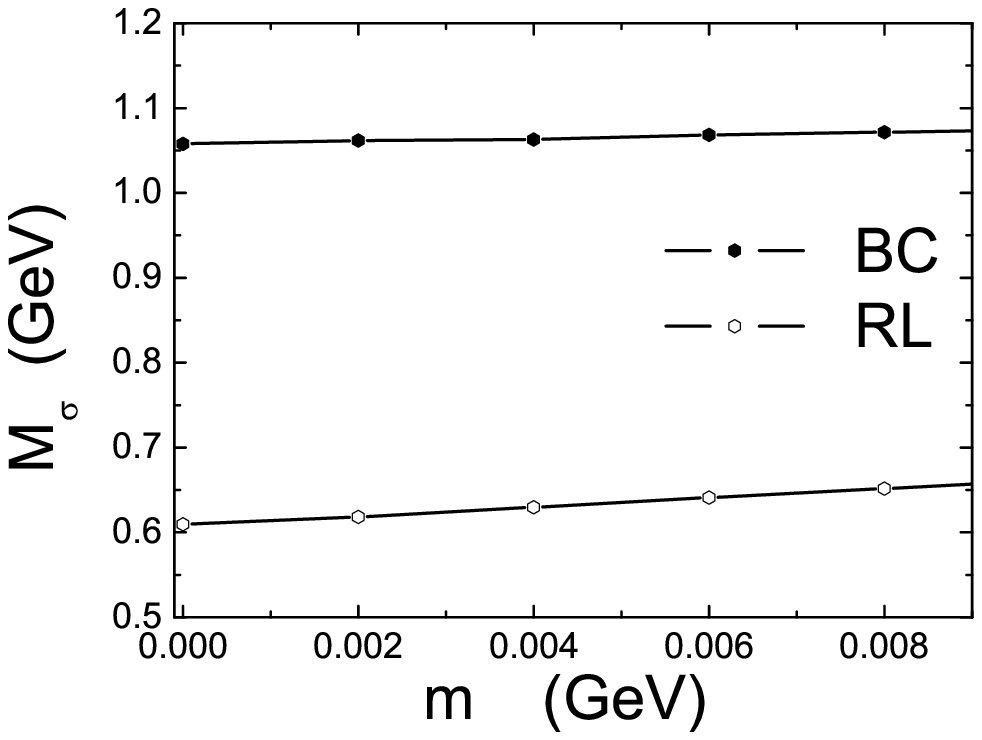}
\vspace*{-5ex}

\caption{\label{massDlarge} Dependence of pseudoscalar (left panel) and scalar (right) meson masses on the current-quark mass, $m$.  The Ball-Chiu vertex (BC) result is compared with the rainbow-ladder (RL) result.  (Figure adapted from Ref.\,\protect\cite{Chang:2009zb}.)}
\end{figure}

A particular novelty of the study is that one can calculate the current-quark-mass-dependence of meson masses using a symmetry-preserving DSE truncation whose diagrammatic content is unknown. That dependence is depicted in Fig.\,\ref{massDlarge} and compared with the rainbow-ladder result.  The $m$-dependence of the pseudoscalar meson's mass provides numerical confirmation of the algebraic fact that the axial-vector Ward-Takahashi identity is preserved by both the rainbow-ladder truncation and the BC-consistent \emph{Ansatz} for the Bethe-Salpeter kernel.  The figure also shows that the axial-vector Ward-Takahashi identity and DCSB conspire to shield the pion's mass from material variation in response to dressing the quark-gluon vertex \cite{Roberts:2007jh,Bhagwat:2004hn}.

As noted in Ref.\,\cite{Chang:2009zb}, a rainbow-ladder kernel with realistic interaction strength yields
\begin{equation}
\label{epsilonRL}
\varepsilon_\sigma^{\rm RL} := \left.\frac{2 M(0) - m_\sigma }{2 M(0)}\right|_{\rm RL} = (0.3 \pm 0.1)\,,
\end{equation}
which can be contrasted with the value obtained using the BC-consistent Bethe-Salpeter kernel; viz.,
\begin{equation}
\label{epsilonBC}
\varepsilon_\sigma^{\rm BC} \lsim 0.1\,.
\end{equation}
Plainly, significant additional repulsion is present in the BC-consistent truncation of the scalar BSE.

Scalar mesons are commonly identified as $^3\!P_0$ states.  This assignment expresses a constituent-quark-model perspective, from which a $J^{PC}=0^{++}$ fermion-antifermion bound-state must have the constituents' spins aligned and one unit of constituent orbital angular momentum; and hence a scalar is a spin and orbital excitation of a pseudoscalar meson.  Of course, no constituent-quark-model can be connected systematically with QCD.  Nevertheless, as we observed above, the presence of orbital angular momentum in a hadron's rest frame is a necessary consequence of Poincar\'e covariance and the vector-boson-exchange character of QCD \cite{Roberts:2007ji,Bhagwat:2006xi,Bhagwat:2006pu}, so there is a realisation in QCD of the quark-model anticipation.

Extant studies of realistic corrections to the rainbow-ladder truncation show that they reduce hyperfine splitting \cite{Bhagwat:2004hn}.  Hence, with the comparison between Eqs.\,(\ref{epsilonRL}) and (\ref{epsilonBC}) one has a clear indication that in a Poincar\'e covariant treatment the BC-consistent truncation magnifies spin-orbit splitting, an effect which can be attributed to the influence of the quark's dynamically-enhanced scalar self-energy \cite{Roberts:2007ji} in the Bethe-Salpeter kernel.

\section{Quark Anomalous Chromomagnetic Moment}
\label{spectrum2}
It was conjectured in Ref.\,\cite{Chang:2009zb} that the full realisation of DCSB in the Bethe-Salpeter kernel will have a material impact on mesons with mass greater than 1\,GeV.  Moreover, that it can overcome a longstanding failure of theoretical hadron physics.  Namely, no extant hadron spectrum calculation is believable because all symmetry preserving studies produce a splitting between vector and axial-vector mesons that is far too small: just one-quarter of the experimental value (see, e.g., Refs.\,\cite{Watson:2004kd,Maris:2006ea,Fischer:2009jm}).

\begin{figure}[t]

\includegraphics[clip,width=0.32\textheight]{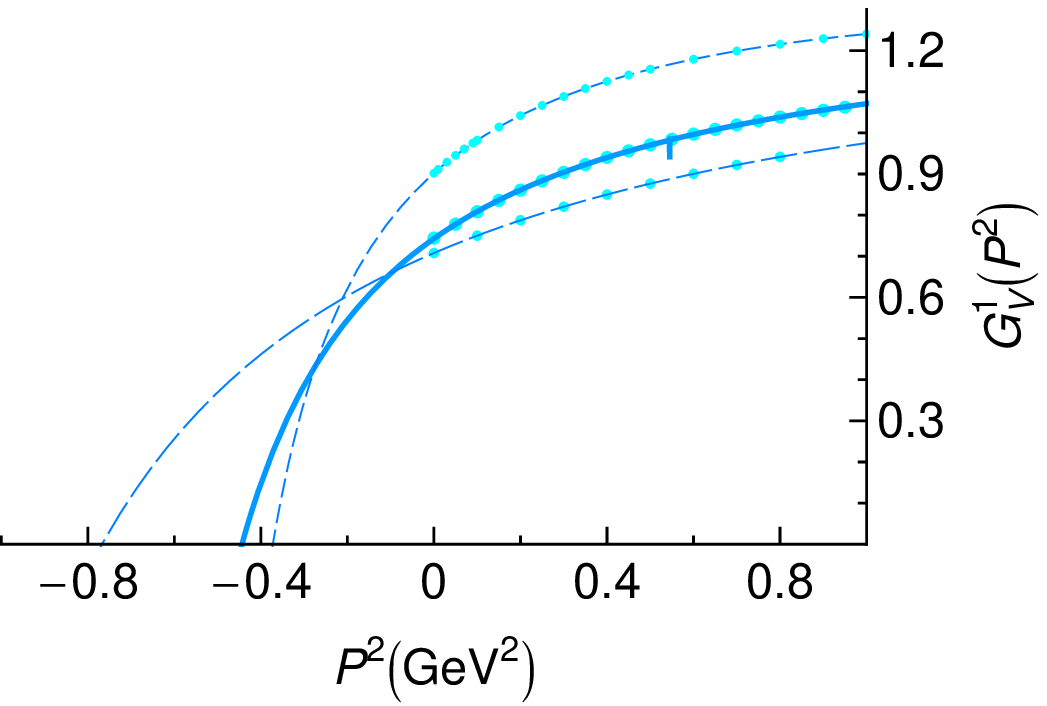}

\includegraphics[clip,width=0.32\textheight]{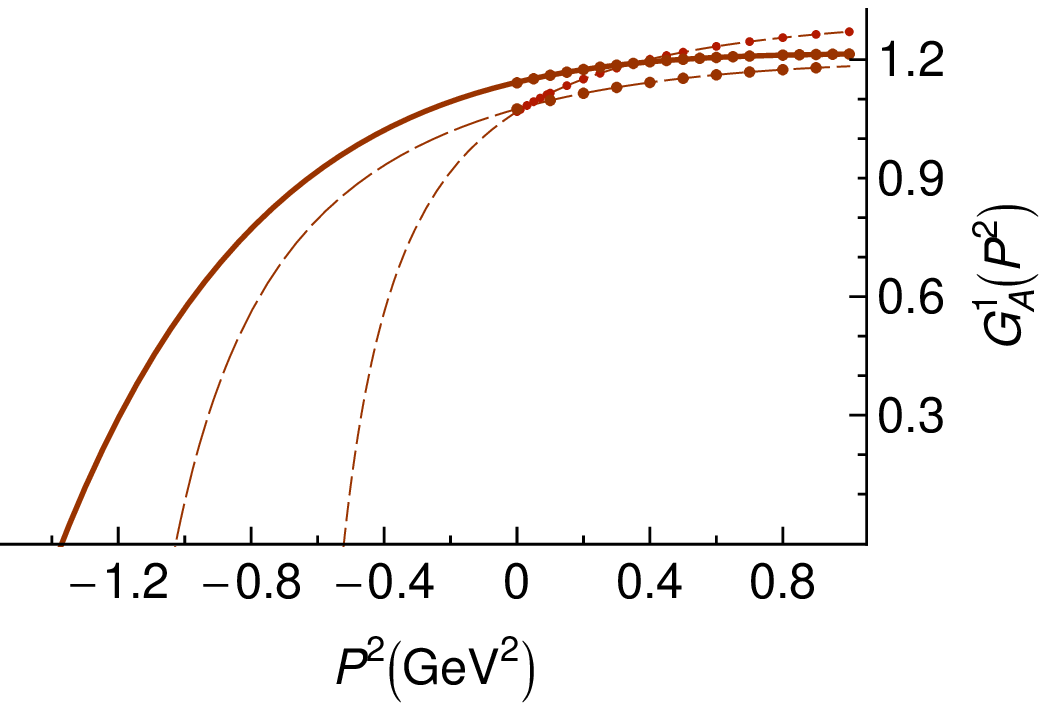}

\caption{\label{a1rho}
\emph{Left panel} -- solution of vector-vertex Bethe-Salpeter equation; and \emph{right panel} -- analogue for the axial-vector vertex.
In both panels, \emph{short-dashed curve} -- rainbow-ladder truncation; \emph{long-dashed curve} -- Ball-Chiu-consistent truncation; and \emph{solid curve} -- improved BC-vertex, with quark anomalous chromomagnetic moment, Eqs.\,(\protect\ref{qcdanom1}), (\protect\ref{qcdanom2}).
From the zero in the curves, one reads the mass-squared of the lightest meson supported by the assumed kernel in each channel.
}
\end{figure}

In this connection, preliminary results\footnote{Preliminary in the sense that not all Dirac covariants are employed in solving for the vertices and internal consistency checks for the pseudoscalar and scalar channels are still underway.} are now available \cite{CR10} and they are summarised in Fig.\,\ref{a1rho}.  The curves are obtained by solving the inhomogeneous vector- and axial-vector-vertex BSEs at spacelike momenta, and fitting a Pad\'e approximant to extrapolate the result to timelike momenta.  Each vertex has a dominant Dirac covariant: $\gamma_\mu \,F_V^1(k;P)$, $\gamma_\mu \,F_A^1(k;P)$, respectively; and the curves in Fig.\,\ref{a1rho} depict the $P^2$-dependence of $1/F^1_{V,A}(k=0;P^2)=:G_{V,A}(P^2)$.  The location of the first zero in such a curve yields the mass-squared of the lightest bound-state produced by the assumed interaction in the associated channel \cite{Bhagwat:2007rj}.  The masses thus determined are listed in Table~\ref{massresult}.

To illustrate the longstanding difficulty, using the interaction described in Ref.\,\cite{Chang:2009zb}, the masses were first calculated in rainbow-ladder truncation; viz., leading-order in the systematic and symmetry-preserving truncation scheme of Ref.\,\cite{Bender:1996bb}.  As anticipated, while the $\rho$-meson mass is acceptable, the $a_1$-mass is far too small.

As we saw above, the procedure introduced in Ref.\,\cite{Chang:2009zb} enables meson masses to be calculated using a symmetry-preserving DSE truncation whose diagrammatic content is unknown.  One can therefore elucidate the effect of an essentially nonperturbative DCSB component in dressed-quark gluon vertex on the $\rho$-$a_1$ complex; in this case, the $\Delta_B$ term in Eq.\,(\ref{bcvtx}) -- recall the enormous impact of this term in the scalar channel, Fig.\,\ref{massDlarge}.  The result is shown in Fig.\,\ref{a1rho}: the DCSB $\Delta_B$-term boosts the $a_1$ mass, which is a positive outcome, but it simultaneously boosts the $\rho$ mass, such that the mass-splitting is practically unchanged from the rainbow-ladder result -- see Table~\ref{massresult}.  Was Ref.\,\cite{Chang:2009zb} too optimistic in expressing a hope that the scheme introduced could provide the first realistic meson spectrum that also encompassed states with mass greater-than 1\,GeV?

Before answering, let us return to a consideration of chirally symmetric QCD.  That theory exhibits helicity conservation so that, perturbatively, the quark-gluon vertex cannot have a term with the helicity-flipping characteristics of $\Delta_B$.  There is another feature of massless fermions in gauge field theories; namely, they cannot posses an anomalous chromo/electro-magnetic moment because the term that describes it couples left- and right-handed fermions.  However, if the theory's chiral symmetry is strongly broken dynamically, why shouldn't the fermions posses a large anomalous chromo/electro-magnetic moment?  Such an effect is expressed in the quark-gluon-vertex via a term
\begin{equation}
\label{qcdanom1}
\Gamma_\mu^{\rm acm} (q,k) = \sigma_{\mu\nu} (q-k)_\nu \, \tau_5(q,k)
\end{equation}
where, owing to DCSB, the natural strength is represented by the \emph{Ansatz}
\begin{equation}
\label{qcdanom2}
\tau_5(q,k) = \Delta_B(q^2,k^2)\,.
\end{equation}
NB.\ Based on the models in Refs.\,\cite{Cloet:2008re,Ivanov:2007cw}, $2 M_E Z(M_E^2) \Delta_B(M_E^2,M_E^2) \sim -\frac{1}{2}$, where $M_E$ is the Euclidean constituent-quark mass, defined in Ref.\,\cite{Maris:1997tm}.

\begin{table}[t]
\begin{tabular}{lcccc}
\hline
  & \tablehead{1}{c}{b}{experiment\\}
  & \tablehead{1}{c}{b}{rainbow-ladder\\}
  & \tablehead{1}{c}{b}{Ball-Chiu consistent\\}
  & \tablehead{1}{c}{b}{Ball-Chiu \\ plus anom.\ cm mom.}\\
  \hline
mass $a_1$ & 1230 & 759 & 1066 & 1230    \\
mass $\rho$& ~775& 644 & ~924 & ~745  \\
mass splitting & ~455 & 115 & ~142 & ~485  \\
\hline
\end{tabular}
\caption{\label{massresult}
Axial-vector and vector meson masses calculated in three truncations of the coupled gap and Bethe-Salpeter equations.  The last column was obtained using the standard Ball-Chiu \emph{Ansatz} augmented by the quark anomalous chromomagnetic moment in Eqs.\,(\protect\ref{qcdanom1}), (\protect\ref{qcdanom2}).}
\end{table}

Using the procedure introduced in Ref.\,\cite{Chang:2009zb}, the vector and axial-vector vertex equations can be solved using the dressed-quark-gluon vertex obtained as the sum of Eqs.\,(\ref{bcvtx}) and (\ref{qcdanom1}).  The effect is remarkable and plain in Fig.\,\ref{a1rho}: the anomalous chromomagnetic moment leads to additional repulsion in the $a_1$ channel but significant attraction in the $\rho$ channel such that, for the first time, a realistic result is simultaneously obtained for the masses in both these channels, and hence the $a_1$-$\rho$ mass-splitting -- see Table~\ref{massresult}.
Furthermore, the origin of the splitting, in an interference between $\Delta_B$ in Eq.\,(\ref{bcvtx}) and $\tau_5$ in Eqs.\,(\protect\ref{qcdanom1}), (\protect\ref{qcdanom2}) is intuitively appealing.  In the chiral limit the mass-splitting between parity partners should owe solely to DCSB and here that is seen explicitly: in the absence of DCSB, $\Delta_B\equiv 0 \equiv \tau_5$.  The rainbow-ladder result is also understood: this truncation fails to adequately express DCSB in the Bethe-Salpeter kernel and hence cannot realistically split parity partners.

Notably, Table~\ref{massresult} is a ``first-guess'' result; i.e., there was no tuning of the strength in Eq.\,(\ref{qcdanom2}), so how reliable can it be?  This question amounts to deciding whether a realistic size is assumed for a light-quark's anomalous chromomagnetic moment.  Fortunately, an analysis is available of results for the dressed-quark-gluon vertex obtained through numerical simulations of quenched-QCD \cite{Skullerud:2003qu}.  This study shows that $\tau_5$ is dynamically two orders-of-magnitude larger than the one-loop perturbative result and, indeed, is of the same magnitude and possesses the same domain of significant support as $\Delta_B(q^2,k^2)$, precisely in accordance with the assumption we have made.

At this point it is natural to consider whether DCSB in QCD can also generate a large quark anomalous \emph{electro}magnetic moment term in the quark-photon vertex.  In perturbation theory, of course, since it doesn't express DCSB, the quark's anomalous electromagnetic moment is small \cite{Bekavac:2009wh}.  One obtains the same answer in the rainbow-ladder truncation; e.g., the $F_6$ and $F_8$ terms in Ref.\,\cite{Maris:1999bh}, which combine to form $\tau_5$ in our notation, contribute less-than 1\% to the pion's electromagnetic form factor.  However, as we've already seen, this truncation doesn't adequately incorporate DCSB into the Bethe-Salpeter kernel.  At present there is no concrete information available that can be used to answer the question but that is about to change.  Upon completion of the study of the $a_1$-$\rho$ complex, we will have determined the magnitude of the analogue of $\tau_5$ in the colour-singlet vector vertex.  It could plausibly be large.

We reiterate that the large quark anomalous chromomagnetic moment owes to DCSB, which itself is probably a consequence of confinement; and real-world electrons and muons do not couple directly to interactions associated with, or degrees-of-freedom affected by, DCSB.  There is thus little reason to anticipate measurable contributions to the electron and muon anomalous electromagnetic moments.  Nevertheless, given the magnitude of the muon ``$g_\mu-2$ anomaly'' and its assumed importance as an harbinger of physics beyond the Standard Model \cite{Bennett:2006fi}, it might be worthwhile to compute a quantitative estimate of the contribution to $a_\mu= (g_\mu-2)/2$ from the quark's DCSB-induced anomalous chromomagnetic, and possibly electromagnetic, moments.  Such contributions will appear in the hadronic component of the photon polarisation tensor.

\section{Pion Electromagnetic Form Factor}
\label{FF1}
In charting the long-range interaction between light-quarks via the feedback between experiment and theory, hadron elastic and transition form factors can provide unique information, beyond that obtained through studies of the hadron spectrum.  This is demonstrated very clearly by an analysis of the electromagnetic pion form factor, $F^{\rm em}_{\pi}(Q^2)$, because the pion has a unique place in the Standard Model.  It is a bound-state of a dressed-quark and -antiquark, and also that almost-massless collective excitation which is the Goldstone mode arising from the dynamical breaking of chiral symmetry.  This dichotomy can only be understood by merging the study of many-body aspects of the QCD vacuum with the symmetry-preserving analysis of two-body bound-states \cite{Maris:1997hd}.  Furthermore, the possibility that this dichotomous nature could have wide-ranging effects on pion properties has made the empirical investigation of these properties highly desirable, despite the difficulty in preparing a system that can act as a pion target and the concomitant complexities in the interpretation of the experiments; e.g., \cite{Volmer:2000ek,Horn:2006tm,Tadevosyan:2007yd,Wijesooriya:2005ir}.

The merit of using $F^{\rm em}_{\pi}(Q^2)$ to elucidate the potential of an interplay between experiment and nonperturbative theory, as a means of constraining the long-range behaviour of QCD's $\beta$-function, is amplified by the existence of a prediction \cite{Farrar:1979aw,Efremov:1979qk,Lepage:1980fj} that $Q^2 F_{\pi}(Q^2)\approx\,$constant for $Q^2\gg m_\pi^2$ in a theory whose interaction is mediated by massless vector-bosons.  The verification of this prediction is a strong motivation for modern experiment \cite{Volmer:2000ek,Horn:2006tm,Tadevosyan:2007yd}, which can also be viewed as an attempt to constrain and map experimentally the pointwise behaviour of the exchange interaction that binds the pion.

Poincar\'e covariance entails that the Bethe-Salpeter amplitude for an isovector pseudoscalar bound-state of a dressed-quark and -antiquark takes the form
\begin{equation}
\Gamma_{\pi}^j(k;P) = \tau^{j}\gamma_5 \left[ i E_\pi(k;P)+ \gamma\cdot P F_\pi(k;P)  + \gamma\cdot P \gamma\cdot k \, G_\pi(k;P) + \sigma_{\mu\nu} k_\mu P_\nu H_\pi(k;P) \right],
\label{genpibsa}
\end{equation}
where $\{\tau^j,j=1,2,3\}$ are the Pauli matrices.  This amplitude is determined from a homogeneous BSE; exemplified, e.g., by Eq.\,(\ref{genbse}) with the driving term, $Z_2 \gamma_5 \gamma_\mu$, omitted.

We can now explicate one of the most important consequences of DCSB.  Namely, a set of four Goldberger-Treiman relations, exact in chiral-limit QCD \cite{Maris:1997hd}:
\begin{equation}
\begin{array}{ll}
\displaystyle f_\pi^0 E_\pi(k;0) = B_0(k^2)\,, &
\displaystyle F_R(k;0) + 2 f_\pi^0 F_\pi(k;0) = A_0(k^2)\,, \\
\displaystyle G_R(k;0) + 2 f_\pi^0 G_\pi(k;0) = A_0^\prime (k^2)\,, &
\displaystyle H_R(k;0) + 2 f_\pi^0 H_\pi(k;0) = 0 \,.
\end{array}
\label{gtpion}
\end{equation}
Here $A_0(k^2)$, $B_0(k^2)$ describe the solution of the chiral limit gap equation; $f_\pi^0$ is the pion's chiral-limit leptonic decay constant; and one has used the following features of the chiral-limit axial-vector vertex:
\begin{eqnarray}
\nonumber
\lefteqn{\Gamma^j_{5\mu}(k;P) \displaystyle \stackrel{P^2\simeq 0}{=}  \frac{P_\mu}{P^2} f_\pi^0 \Gamma^j_\pi(k;P)} \\
&& + \frac{\tau^{j}}{2}\gamma_5 \left[ \gamma_\mu F_\pi(k;P)  + \gamma_\mu \gamma\cdot k \, G_\pi(k;P) - \sigma_{\mu\nu} k_\nu H_\pi(k;P) \right] + \Gamma_{5\mu}^{\rm reg}(k;P)\,,
\label{genavbsa}
\end{eqnarray}
where $F_R$, $G_R$, $H_R$, $\Gamma_{5\mu}^{\rm reg}(k;P)$ are regular for $P^2 \simeq 0$ and $P_\mu\Gamma_{5\mu}^{\rm reg}(k;P)\sim\,$O$(P^2)$ for $P^2 \simeq 0$.

Equations~(\ref{gtpion}) are fascinating and have far-reaching consequences.  The first identity states that, owing to DCSB, the dominant piece of the pseudoscalar meson Bethe-Salpeter amplitude is completely determined by the scalar piece of the dressed-quark self-energy; i.e., by solving the quark one-body problem, one obtains simultaneously the solution for an important part of the pseudoscalar meson two-body problem.  The next two demonstrate that a pseudoscalar meson necessarily contains components of pseudovector origin.  These terms alter the asymptotic form of $F_{\pi}^{\rm em}(Q^2)$ by a factor of $Q^2$ cf.\ the result obtained in their absence \cite{Maris:1998hc}.  One may consider Eqs.\,(\ref{gtpion}) as an expression of pointwise consequences of Goldstone's theorem, which is therefore seen to have an impact far broader than that described in textbooks.

QCD-based DSE calculations of $F^{\rm em}_\pi(Q^2)$ exist \cite{Maris:1998hc,Maris:2000sk}, the most systematic of which \cite{Maris:2000sk} predicted the measured form factor \cite{Volmer:2000ek}.  Germane to our discourse, however, is an elucidation of the sensitivity of $F^{\rm em}_\pi(Q^2)$ to the pointwise behaviour of the interaction between quarks.  We therefore recapitulate on Ref.\,\cite{GutierrezGuerrero:2010md}, which explored how predictions for pion properties change if quarks interact not via massless vector-boson exchange but instead through a contact interaction; viz.,
\begin{equation}
\label{njlgluon}
g^2 D_{\mu \nu}(p-q) = \delta_{\mu \nu} \frac{1}{m_G^2}\,,
\end{equation}
where $m_G$ is a gluon mass-scale (such a scale is generated dynamically in QCD, with a value $\sim 0.5\,$GeV \cite{Bowman:2004jm}), and proceeded by embedding this interaction in a rainbow-ladder truncation of the DSEs.

In this case, using a confining regularisation scheme \cite{Ebert:1996vx}, the gap equation, which determines this interaction's momentum-independent dressed-quark mass, can be written
\begin{equation}
M = m +  \frac{M}{3\pi^2 m_G^2} \,{\cal C}(M^2;\tau_{\rm ir},\tau_{\rm uv})\,,
\end{equation}
where $m$ is the current-quark mass and ${\cal C}/M^2 = \Gamma(-1,M^2 \tau_{\rm uv}^2) - \Gamma(-1,M^2 \tau_{\rm ir}^2)$, with $\Gamma(\alpha,y)$ being the incomplete gamma-function.  Results are presented in Table\,\ref{Table:Para1}.

\begin{table}[t]
\begin{tabular}{cccccccccc}
\hline
    \tablehead{1}{c}{b}{${\cal N}$}
& \tablehead{1}{c}{b}{$E_\pi^{\rm c}$}
& \tablehead{1}{c}{b}{$F_\pi^{\rm c}$}
& \tablehead{1}{c}{b}{$F_R$}
& \tablehead{1}{c}{b}{$M$}
& \tablehead{1}{c}{b}{$\kappa$}
& \tablehead{1}{c}{b}{$f_\pi^0$}
& \tablehead{1}{c}{b}{$\displaystyle\left.f_\pi^0\right|_{F_\pi\to 0}$}
& \tablehead{1}{c}{b}{$r_\pi^0$}
& \tablehead{1}{c}{b}{$\displaystyle\left.r_\pi^0\right|_{F_\pi\to 0}$}\\
\hline
0.23 & 4.28 & 0.69 & 0.68 & 0.40 & 0.22 & 0.094 & 0.11 & 0.29 & 0.41\\
\end{tabular} 
\caption{Results obtained with (in GeV) $m=0$, $m_G=0.11\,$, $\Lambda_{\rm ir} = 1/\tau_{\rm ir} = 0.24\,$, $\Lambda_{\rm uv}=1/\tau_{\rm uv} = 0.823$.  They are commensurate with those from QCD-based DSE studies \protect\cite{Maris:1997tm}.  Dimensioned quantities are listed in GeV or fm, as appropriate, and $\kappa := -\langle \bar q q \rangle^{1/3}$.
\label{Table:Para1}
}
\end{table}

With a symmetry-preserving regularisation of the interaction in Eq.\,(\ref{njlgluon}), the Bethe-Salpeter amplitude cannot depend on relative momentum.  Hence Eq.\,(\ref{genpibsa}) becomes
\begin{equation}
\Gamma_\pi(P) = \gamma_5 \left[ i E_\pi(P) + \frac{1}{M} \gamma\cdot P F_\pi(P) \right]
\end{equation}
and the explicit form of the model's ladder BSE is
\begin{equation}
\label{bsefinal0}
\left[
\begin{array}{c}
E_\pi(P)\\
F_\pi(P)
\end{array}
\right]
= \frac{1}{3\pi^2 m_G^2}
\left[
\begin{array}{cc}
{\cal K}_{EE} & {\cal K}_{EF} \\
{\cal K}_{FE} & {\cal K}_{FF}
\end{array}\right]
\left[\begin{array}{c}
E_\pi(P)\\
F_\pi(P)
\end{array}
\right],
\end{equation}
where, with $m=0=P^2$, anticipating the Goldstone character of the pion,
\begin{equation}
\begin{array}{cl}
{\cal K}_{EE}  =  {\cal C}(M^2;\tau_{\rm ir}^2,\tau_{\rm uv}^2)\,, &  {\cal K}_{EF}  =  0\,,\\
2 {\cal K}_{FE} = {\cal C}_1(M^2;\tau_{\rm ir}^2,\tau_{\rm uv}^2) \,,&
{\cal K}_{FF} = - 2 {\cal K}_{FE}\,,
\end{array}
\end{equation}
and ${\cal C}_1(z) = - z {\cal C}^\prime(z)$, where we have suppressed the dependence on $\tau_{\rm ir,uv}$.  The solution of Eq.\,(\ref{bsefinal0}) gives the pion's chiral-limit Bethe-Salpeter amplitude:
\begin{equation}
\label{EpiFpi1}
E^1_\pi = 0.987 \,,\; F^1_\pi=0.160\,,
\end{equation}
written with unit normalisation.

However, this is not the physical convention.  The canonical normalisation procedure ensures unit residue for the pion bound-state contribution to the quark-antiquark scattering matrix, a property of
$\Gamma_\pi^{\rm c}(P) =\frac{1}{{\cal N}} \, \Gamma_\pi^1(P)$, where
\begin{equation}
%
{\cal N}^2 P_\mu = N_c\, {\rm tr} \int\! \frac{d^4q}{(2\pi)^4}\Gamma_\pi^1(-P)
 \frac{\partial}{\partial P_\mu} S(q+P) \, \Gamma_\pi^1(P)\, S(q)\,. \label{Ndef}
\end{equation}
In the chiral limit,
\begin{equation}
{\cal N}_0^2 = \frac{N_c}{4\pi^2} \frac{1}{M^2} \, {\cal C}_1(M^2;\tau_{\rm ir}^2,\tau_{\rm uv}^2)
E_\pi^1 [ E_\pi^1 - 2 F_\pi^1].
\label{Norm0}
\end{equation}
The pion's leptonic decay constant is obtained from the canonically normalised amplitude
and in the chiral limit
\begin{equation}
\label{fpi0}
f_\pi^0 = \frac{N_c}{4\pi^2} \frac{1}{M} {\cal C}_1(M^2;\tau_{\rm ir}^2,\tau_{\rm uv}^2)  [ E_\pi^{\rm c} - 2 F_\pi^{\rm c} ]\,.
\end{equation}

If one has preserved Eq.\,(\ref{avwtim}), then, for $m=0$ in the neighbourhood of $P^2=0$, the solution of the axial-vector BSE has the form -- cf.\, Eq.\,(\ref{genavbsa}):
\begin{equation}
\Gamma_{5\mu}(k_+,k) = \frac{P_\mu}{P^2} \, 2 f_\pi^0 \, \Gamma^{\rm c}_\pi(P) + \gamma_5\gamma_\mu F_R(P)
\end{equation}
and the following subset of Eqs.\,(\ref{gtpion}) will hold:
\begin{equation}
\label{GTI}
f_\pi^0 E^{\rm c}_\pi = M \,,\; 2\frac{F^{\rm c}_\pi}{E^{\rm c}_\pi}+ F_R = 1\,.
\end{equation}
That they do can be verified from Table\,\ref{Table:Para1}, which also shows that $F_\pi(P)$, necessarily nonzero in a vector exchange theory, irrespective of the pointwise behaviour of the interaction, has a measurable impact on the value of $f_\pi$.

Based upon these results, one can proceed to compute the electromagnetic pion form factor in the generalised impulse approximation \cite{GutierrezGuerrero:2010md}; i.e., at leading-order in a symmetry-preserving DSE truncation scheme \cite{Maris:1998hc,Maris:2000sk,Roberts:1994hh}. Namely, for an incoming pion with momentum $p_1=K-Q/2$, which absorbs a photon with space-like momentum $Q$, so that the outgoing pion has momentum $p_2=K+Q/2$,
\begin{equation}
2 K_{\mu} F_{\pi}^{\rm em}(Q^2) = 2 N_c \int\frac{d^4t}{(2\pi)^4}
{\rm tr_D} \Bigg[ i \Gamma_{\pi}^{\rm c}(-p_2) S(t+p_2) i \gamma_{\mu}  S(t+p_1) \; i \Gamma_{\pi}^{\rm c}(p_1) \; S(t) \Bigg].
\label{KF}
\end{equation}
The form factor is expressible as a sum; viz.,
\begin{eqnarray}
F_{\pi}^{\rm em}(Q^2) &=& F_{\pi,EE}^{{\rm em}}(Q^2) + F_{\pi,EF}^{{\rm em}}(Q^2) + F_{\pi,FF}^{{\rm em}}(Q^2),\\
& =& E_\pi^{\rm c}\,\!^2  T^{\pi}_{EE}(Q^2) + E_\pi^{\rm c} F_\pi^{\rm c} T^{\pi}_{EF}(Q^2) + F_\pi^{\rm c}\,\!^2 T^{\pi}_{FF}(Q^2) ,
\label{F123}
\end{eqnarray}
wherein each function $T^{\pi}$ has a simple algebraic form in this model.

\begin{figure}[t] 
\includegraphics[clip,height=0.34\textheight,angle=-90]{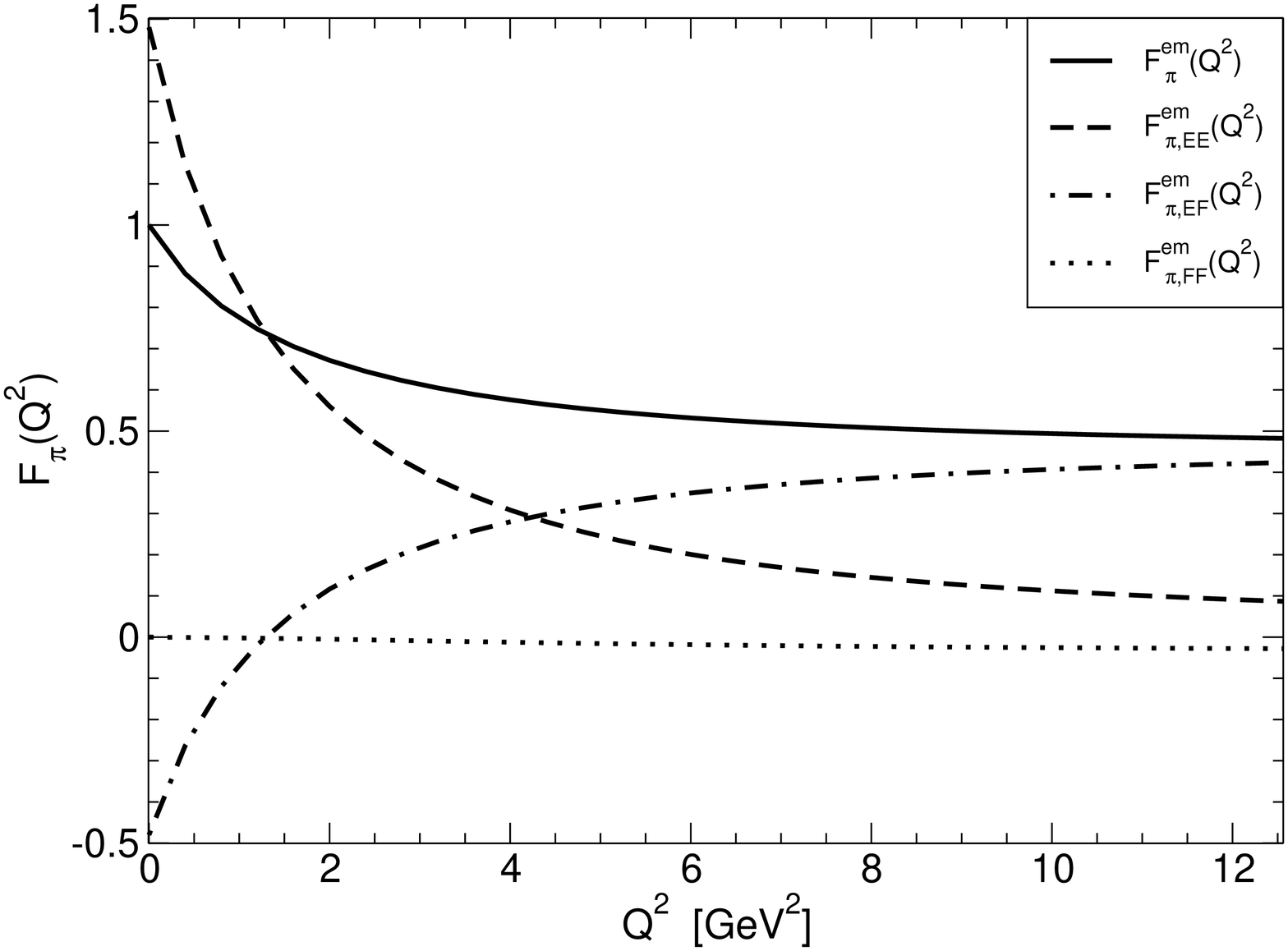}\hspace*{0.3em}
\includegraphics[clip,height=0.34\textheight,angle=-90]{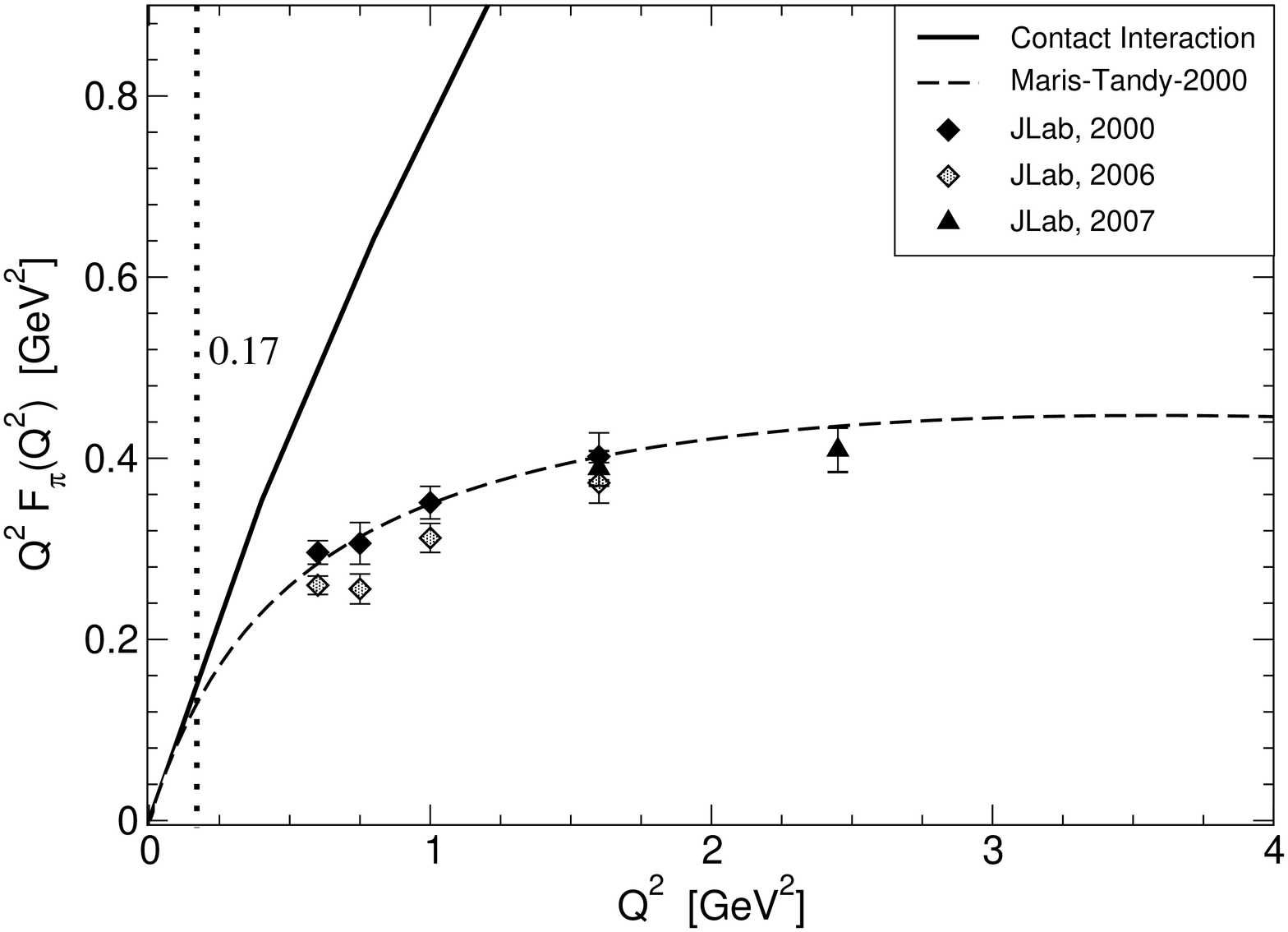}
\caption{\label{fig4}
\underline{Left panel}.
$F^{\rm em}_{\pi}(Q^2)$ and the separate contributions introduced in Eq.\,(\protect\ref{F123}).
$F^{\rm em}_{\pi}(Q^2=0)=1$, without fine-tuning, because a symmetry-preserving regularisation of the interaction in Eq.\,(\protect\ref{njlgluon}) was implemented.
\underline{Right panel}.
\emph{Solid curve}: $Q^2 F_{\pi}(Q^2)$ obtained with Eq.\,(\protect\ref{njlgluon}).
%
\emph{Dashed curve}: DSE prediction \protect\cite{Maris:2000sk}, which employed a momentum-dependent renormalisation-group-improved gluon exchange interaction.
For $Q^2>0.17\,$GeV$^2\approx M^2$, marked by the vertical \emph{dotted line}, the contact interaction result for $F^{\rm em}_{\pi}(Q^2)$ differs from that in  Ref.\,\protect\cite{Maris:2000sk} by more than 20\%.  The data are from Refs.\,\protect\cite{Volmer:2000ek,Horn:2006tm,Tadevosyan:2007yd}.
(Figure adapted from Ref.\,\cite{GutierrezGuerrero:2010md}.)
}
\end{figure}

In the left panel of Fig.\,\ref{fig4} we present $F_\pi^{\rm em}(Q^2)$ and the three separate contributions defined in Eq.\,(\ref{F123}).
We highlight two features.  First, $F_{\pi,EF}^{\rm em}(Q^2=0)$ contributes roughly one-third of the pion's unit charge.  This could have been anticipated from Eq.\,(\ref{Norm0}).
Second, and perhaps more dramatic: the interaction in Eq.\,(\ref{njlgluon}) generates \begin{equation}
\label{Fpiconstant}
F_\pi^{\rm em}(Q^2 \to\infty) =\,{\rm constant.}
\end{equation}
Both results originate in the nonzero value of $F_\pi(P)$, which is a straightforward consequence of the symmetry-preserving treatment of a vector exchange theory \cite{Maris:1997hd}.  Equation~(\ref{Fpiconstant}) should not come as a surprise: with a symmetry-preserving regularisation of the interaction in Eq.\,(\ref{njlgluon}), the pion's Bethe-Salpeter amplitude cannot depend on the constituent's relative momentum.  This is characteristic of a pointlike particle, which must have a hard form factor.

A striking outcome, evident in Fig.\,\ref{fig4}: $F_{\pi,EE}^{\rm em}$ is only a good approximation to the net pion form factor for $Q^2 \lsim M^2$.  $F_{\pi,EE}^{\rm em}$ and $F_{\pi,EF}^{\rm em}$ evolve with equal rapidity -- there is no reason for this to be otherwise, as they are determined by the same mass-scales -- but a nonzero constant comes quickly to dominate over a form factor that falls swiftly to zero.

In the right panel of Fig.\,\ref{fig4} we compare the form factor computed from Eq.\,(\ref{njlgluon}) with contemporary experimental data \cite{Volmer:2000ek,Horn:2006tm,Tadevosyan:2007yd} and a QCD-based DSE prediction \cite{Maris:2000sk}.  Both the QCD-based result and that obtained from the momentum-independent interaction yield the same values for the pion's static properties. However, for $Q^2>0$ the form factor computed using $\sim 1/k^2$ vector boson exchange is immediately distinguishable empirically from that produced by a momentum-independent interaction.  Indeed, the figure shows that for $F_\pi^{\rm em}$, existing experiments can already distinguish between different possibilities for the quark-quark interaction.

This summary illustrates that when a momentum-independent vector-exchange interaction is regularised in a symmetry-preserving manner, the results are directly comparable with experiment, computations based on well-defined and systematically-improvable truncations of QCD's DSEs \cite{Maris:2000sk}, and perturbative QCD.  In this context it will now be apparent that a contact interaction, whilst capable of describing pion static properties well, Table\,\ref{Table:Para1}, generates a form factor whose evolution with $Q^2$ deviates markedly from experiment for $Q^2>0.17\,$GeV$^2\approx M^2$ and produces asymptotic power-law behaviour, Eq.\,(\ref{Fpiconstant}), in serious conflict with perturbative-QCD \cite{Farrar:1979aw,Efremov:1979qk,Lepage:1980fj}.

In closing this section we note that the contact interaction produces a momentum-independent dressed-quark mass function, in contrast to QCD-based DSE studies \cite{Roberts:2007ji,Bhagwat:2006tu} and lattice-QCD \cite{Bowman:2005vx}.  This is fundamentally the origin of the marked discrepancy between the form factor it produces and extant experiment.  Hence Ref.\,\cite{GutierrezGuerrero:2010md} highlights that form factor observables, measured at an upgraded Jefferson laboratory, e.g., are capable of mapping the running of the dressed-quark mass function.  Efforts are underway to establish the signals of the running mass in baryon elastic and transition form factors.

\section{Pion and kaon valence-quark distributions}
\label{FF2}
The past forty years has seen a tremendous effort to deduce the parton distribution
functions of the most accessible hadrons -- the proton, neutron and pion.  There are many reasons for this long sustained and thriving interest \cite{Holt:2010vj} but in large part it is motivated by the suspected process-independence of the usual parton distribution functions and hence an ability to unify many hadronic processes through their computation.  In connection with uncovering the essence of the strong interaction, the behaviour of the valence-quark distribution functions at large Bjorken-$x$ is most relevant.  Importantly, in the infinite momentum frame, Bjorken-$x$ measures the fraction of a hadron's four-momentum carried by the struck parton and, e.g., the valence-quark distribution function, $q_{\rm v}(x)$, measures the number-density of valence-quarks with momentum-fraction $x$.

Owing to the dichotomous nature of Goldstone bosons, understanding the valence-quark distribution functions in the pion and kaon is of great importance.  Moreover, given the large value of the ratio of $s$-to-$u$ current-quark masses, a comparison between the pion and kaon structure functions offers the chance to chart effects of explicit chiral symmetry breaking on the structure of would-be Goldstone modes.  There is also the prediction \cite{Ezawa:1974wm,Farrar:1975yb} that a theory in which the quarks interact via $1/k^2$ vector-boson exchange will produce valence-quark distribution functions for which
\begin{equation}
q_{\rm v}(x) \propto (1-x)^{2+\gamma} \,,\; x\gsim 0.85\,,
\end{equation}
where $\gamma\gsim 0$ is an anomalous dimension that grows with increasing momentum transfer.  (See Sec.VI.B.3 of Ref.\,\cite{Holt:2010vj} for a detailed discussion.)

Experimental knowledge of the parton structure of the pion and kaon arises primarily from pionic or kaonic Drell-Yan scattering from nucleons in heavy nuclei \cite{Wijesooriya:2005ir,Badier:1980jq,Conway:1989fs}.  Theoretically, given that DCSB plays a crucial role in connection with pseudoscalar mesons, one must employ an approach that realistically expresses this phenomenon.  The DSEs therefore provide a natural framework: a study of the pion exists \cite{Hecht:2000xa} and one of the kaon is underway \cite{kaonpdf}.

One can illustrate the results anticipated from the latter study through an internally consistent calculation based upon the QCD-improvement of a simple model used already for pion and kaon distribution functions \cite{Shigetani:1993dx}.  In its original form, as a version of the Nambu-Jona--Lasinio model with a hard-cutoff, $\Lambda=0.9\,$GeV, and momentum-independent dressed-quark masses, $\check M_u=0.35\,$GeV and $\check M_s=0.53\,$GeV, the model yields valence-quark distribution functions that behave as $(1-x)$ at large $x$, in conflict with perturbative QCD and nonperturbative DSE studies.  However, this can be remedied through the inclusion of a pion Bethe-Salpeter amplitude whose dependence on relative momentum is QCD-like; viz., $\phi_\pi(k^2)\propto 1/k^2$ when the magnitude of $k^2$ is large.
%
To be explicit,
\begin{eqnarray}
\label{uvSphi}
u_{\rm v}^{\pi}(x) & = & -g_{\pi\bar d u}^2 \int_{\kappa_{\rm ir}}^{\kappa_{\rm uv}} d\kappa
\frac{ \kappa - M_u(\kappa)^2 - x m_\pi^2}{(\kappa - M_u(\kappa)^2)^2} \phi_\pi(\kappa)\,,
\end{eqnarray}
where $\kappa_{\rm ir}= - x (\check M_u^2 - m_\pi^2 (1-x))/(1-x)$, with $\kappa_{\rm ir}-\kappa_{\rm uv}=\Lambda^2$;
\begin{equation}
\phi_\pi(\kappa) = \frac{\exp(\kappa/\Lambda_\pi^2) - 1}{\kappa/\Lambda_\pi^2}\,;\;
M_u(\kappa) = m_u + (\check M_u - m_u) \phi_\pi(\kappa)\,,
\end{equation}
which is an \emph{Ansatz} for the momentum-dependent dressed-quark mass, $M_u(k^2)$, with $m_u=5.5\,$MeV, in a form motivated by the first of Eqs.\,(\ref{gtpion}); and the value of the pion-quark-antiquark coupling constant, $g_{\pi\bar d u}$, ensures
\begin{equation}
\label{uvpinorm}
\int_0^1 dx\; u_{\rm v}^\pi(x) = 1\,.
\end{equation}
A value of $\Lambda_\pi=0.6\,$GeV provides a satisfactory least-squares fit to $\{u_{\rm v}^\pi(x), x>0.4\}$ in Ref.\,\cite{Hecht:2000xa}.  (NB.\ The hard cutoff in Eq.\,(\ref{uvSphi}), mimics the effect of a second Bethe-Salpeter amplitude, which would otherwise appear.)

Repeating the analysis for the kaon, we obtain
\begin{eqnarray}
u_{\rm v}^{K}(x) & = & -g_{K\bar s u}^2 \int_{\tilde\kappa_{\rm ir}}^{\tilde\kappa_{\rm uv}} d\kappa
\frac{ \kappa - M_u(\kappa)^2 + x (\check M_u^2-\check M_s^2-m_K^2)}
{(\kappa - M_u(\kappa)^2)^2} \phi_K(\kappa)\,,
\end{eqnarray}
where $\tilde\kappa_{\rm ir}= - x (\check M_s^2 - m_K^2 (1-x))/(1-x)$, with $\tilde \kappa_{\rm ir}-\tilde\kappa_{\rm uv}=\Lambda^2$;
\begin{equation}
\phi_\pi(\kappa) = \frac{\exp(\kappa/\Lambda_K^2) - 1}{\kappa/\Lambda_K^2}\,,\;
M_s(\kappa) = m_s + (M_s - m_s) \phi_K(\kappa)\,,
\end{equation}
with $m_s=0.135\,$GeV and $\Lambda_K=1.1 \Lambda_\pi$, the latter value motivated by the BSE results in Ref.\,\cite{Maris:1997tm}; and the value of $g_{K\bar s u}$ ensures the kaon analogue of Eq.\,(\ref{uvpinorm}).

\begin{table}[t]
\begin{tabular}{clllll}
\hline
  \tablehead{1}{c}{b}{$\,$}
& \tablehead{1}{c}{b}{$u_{\rm v}^\pi(x;Q_0)$ \protect\cite{Hecht:2000xa}}
& \tablehead{1}{c}{b}{$u_{\rm v}^\pi(x;Q_0)$}
& \tablehead{1}{c}{b}{$u_{\rm v}^\pi(x;Q)$}
& \tablehead{1}{c}{b}{$u_{\rm v}^K(x;Q_0)$}
& \tablehead{1}{c}{b}{$u_{\rm v}^K(x;Q)$}\\
\hline
$\langle x \rangle$     & 0.36 & 0.35 & 0.21  & 0.32  & 0.19  \\
$\langle x^2 \rangle$   & 0.18 & 0.17 & 0.078 & 0.15  & 0.066 \\
$\langle x^3 \rangle$   & 0.10 & 0.10 & 0.037 & 0.078 & 0.029 \\\hline
\end{tabular}
\caption{Leading moments of our valence $u$-quark distributions: $\langle x^n\rangle_{Q} := \int_0^1 dx\; x^n u_{\rm v}(x;Q)$, with $Q_0=0.57\,$GeV and $Q=5\,$GeV.  For comparison, the first numerical column shows results from Ref.\,\protect\cite{Hecht:2000xa}.
\label{Table:moments}
}
\end{table}

In Table~\ref{Table:moments} we report low moments of the computed valence-quark distribution functions.  The near numerical agreement between entries in the first two columns is an illustration of the similarity between our model distribution and that by which it was constrained \cite{Hecht:2000xa}.  However, it should be borne in mind that these low moments are not very sensitive to the pointwise behaviour of distributions near $x=1$ -- see, e.g., Secs.\,VI.B.2 and VI.B.3 of Ref.\,\protect\cite{Holt:2010vj}.

\begin{figure}[t]
\vspace*{-7ex}

\includegraphics[clip,width=0.33\textheight]{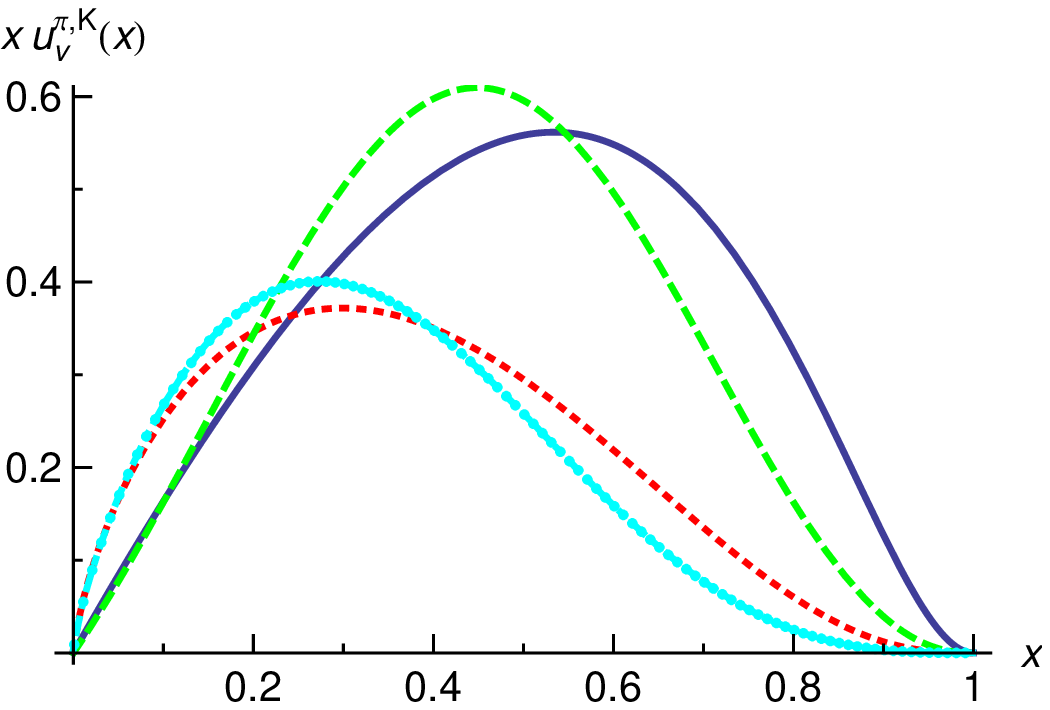}\hspace*{0.5em}
\vspace*{-7ex}

\includegraphics[clip,width=0.33\textheight]{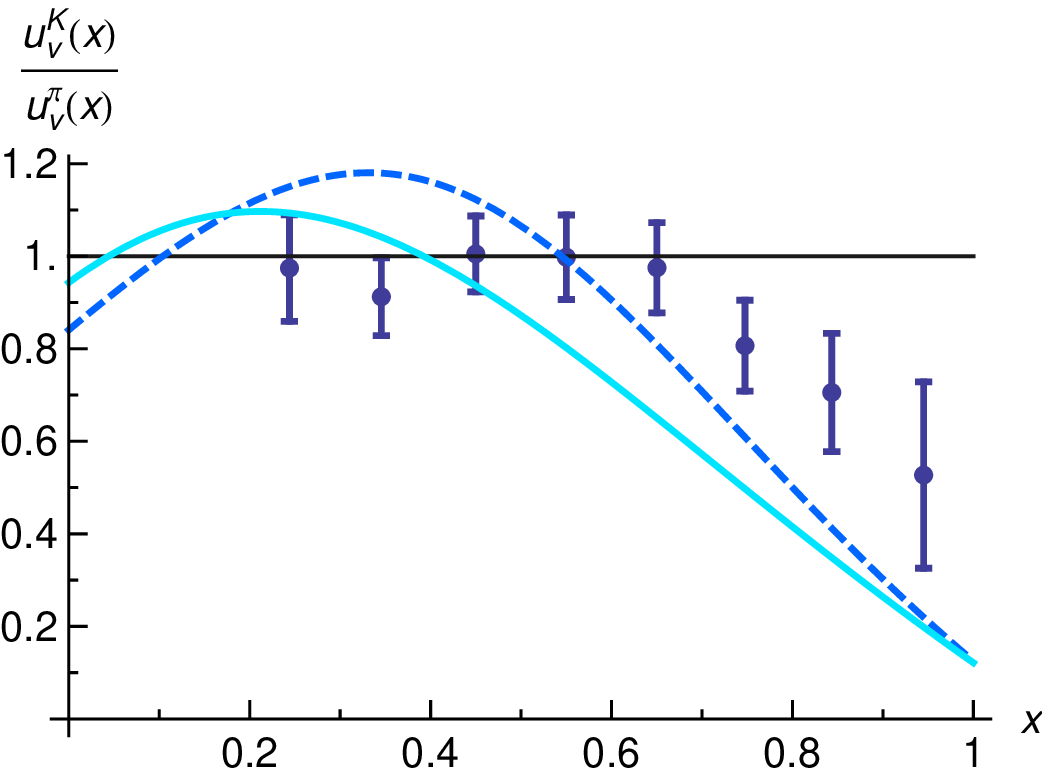}
\vspace*{-5ex}

\caption{\label{ShigetaniSoft}
\underline{Left panel}. Valence $u$-quark distribution functions, computed as described in the text: \emph{solid curve}, $u_{\rm v}^\pi(x)$; and \emph{dashed curve}, $u_{\rm v}^K(x)$.  Applying leading-order QCD evolution from $Q_0^2=0.32\,$GeV$^2$ to $Q^2=25\,$GeV$^2$, explained in Sec.\,II.D of Ref.\,\protect\cite{Holt:2010vj}, one obtains the other two curves from these starting distributions: \emph{dashed curve}, $u_{\rm v}^\pi(x;25)$; and \emph{dotted curve}, $u_{\rm v}^K(x,25)$.
\underline{Right panel}.  Model ratio $u_{\rm v}^K/u_{\rm v}^\pi$ evaluated at $Q_0^2=0.32\,$GeV$^2$, \emph{Dashed curve}, and $Q^2=25\,$GeV$^2$, \emph{solid curve}. $\left. u_{\rm v}^K/u_{\rm v}^\pi\right|_{x=1}=0.13$.  Under the right conditions, $u_{\rm v}^K/u_{\rm v}^\pi$ should equal the ratio of kaon-to-pion Drell-Yan cross-sections, and we reproduce that obtained from a sample of dimuon events with invariant mass $4.1<M<8.5\,$GeV \protect\cite{Badier:1980jq}.}
\end{figure}

In Fig.\,\ref{ShigetaniSoft} we depict our computed distributions themselves and relevant ratios.  Aspects of the curves are model-independent.
For example, owing to its larger mass, one anticipates that the $s$-quark should carry more of the charged-kaon's momentum than the $u$-quark.  This explains, in a manner which is transparent in this model, why the support of $x u_{\rm v}^K(x)$ is shifted to lower-$x$ than that $x u_{\rm v}^\pi(x)$.
QCD evolution is an area-conserving operation on the distribution function, which shifts support from large-$x$ to small-$x$.  Thus, while both $u_{\rm v}^{\pi,K}(x;Q_0) \propto (1-x)^2$ for $x\simeq 1$,
\begin{equation}
u_{\rm v}^{\pi,K}(x;Q) \stackrel{x\simeq 1}{\propto} (1-x)^a, \; a= 2.7\,.
\end{equation}
These observations explain the qualitative behaviour of the evolved distributions.
Concerning the ratio, as a consequence of the form of the evolution equations, the value of the ratio at $x=1$ is invariant under evolution.  At the other extreme, the value of the ratio at $x=0$ approaches one under evolution owing to the increasingly large population of sea-quarks produced thereby.

\section{Baryon Properties}
\label{FF3}
While a symmetry-preserving description of mesons is essential, it is only part of the physics that nonperturbative QCD must describe since Nature also presents baryons, light-quarks in three-particle composites.  An explanation of the spectrum of baryons and the nature of interactions between them is basic to understanding the Standard Model.  The present and planned experimental programmes at JLab, and other facilities worldwide, are critical elements in this effort.

No approach to QCD is comprehensive if it cannot provide a unified explanation of both mesons and baryons.  We have explained that DCSB is a keystone of the Standard Model, which is evident in the momentum-dependence of the dressed-quark mass function -- Fig.\,\ref{gluoncloud}: it is just as important to baryons as it is to mesons.  The DSEs furnish the only extant framework that can simultaneously connect both meson and baryon observables with this basic feature of QCD, having provided, e.g., a direct correlation of meson and baryon properties via a single interaction kernel, which preserves QCD's one-loop renormalisation group behaviour and can systematically be improved \cite{Eichmann:2008ae,Eichmann:2008ef}.

In quantum field theory a baryon appears as a pole in a six-point quark Green function.  The residue is proportional to the baryon's Faddeev amplitude, which is obtained from a Poincar\'e covariant Faddeev equation that sums all possible exchanges and interactions that can take place between three dressed-quarks.  A tractable Faddeev equation for baryons \cite{Cahill:1988dx} is founded on the observation that an interaction which describes colour-singlet mesons also generates nonpointlike quark-quark (diquark) correlations in the colour-$\bar 3$ (antitriplet) channel \cite{Cahill:1987qr}.  The lightest diquark correlations appear in the $J^P=0^+,1^+$ channels \cite{Burden:1996nh,Maris:2002yu} and hence only they need be retained in approximating the quark-quark scattering matrix that appears as part of the Faddeev equation \cite{Cloet:2008re,Eichmann:2008ef}.

Diquarks do not appear in the strong interaction spectrum \cite{Bender:1996bb,Bhagwat:2004hn} but the attraction between quarks in this channel justifies a picture of baryons in which two quarks within a baryon are always correlated as a colour-$\bar 3$ diquark pseudoparticle, and binding is effected by the iterated exchange of roles between the bystander and diquark-participant quarks.   Here it is important to emphasise strongly that QCD supports \emph{nonpointlike} diquark correlations \cite{Maris:2004bp}.  Hence models that employ pointlike diquark degrees of freedom have little connection with QCD.

Numerous properties of the nucleon have been computed using the Faddeev equation just described. Herein we draw only two examples from comprehensive analyses of nucleon electromagnetic form factors \cite{Cloet:2008re,Chang:2009ae,Cloet:2008wg,Cloet:2008fw}.  To introduce the results, we note that the nucleon's electromagnetic current is
\begin{equation}
\label{Jnucleon}
J_\mu(P^\prime,P) =
  i e \,\bar u(P^\prime)\,\left( \gamma_\mu F_1(Q^2) +
\frac{1}{2M}\, \sigma_{\mu\nu}\,Q_\nu\,F_2(Q^2)\right) u(P)\,,
\end{equation}
where $P$ ($P^\prime$) is the momentum of the incoming (outgoing) nucleon, $Q= P^\prime - P$ is the momentum transfer, and $F_1$ and $F_2$ are, respectively, the Dirac and Pauli form factors,
from which one obtains the nucleon's electric and magnetic (Sachs) form factors
\begin{equation}
\label{GEpeq}
G_E(Q^2)  =  F_1(Q^2) - \frac{Q^2}{4 M^2} F_2(Q^2)\,,\;
G_M(Q^2)  =  F_1(Q^2) + F_2(Q^2)\,.
\end{equation}

\begin{figure}[t]
\vspace*{-7ex}

\includegraphics[clip,width=0.30\textheight]{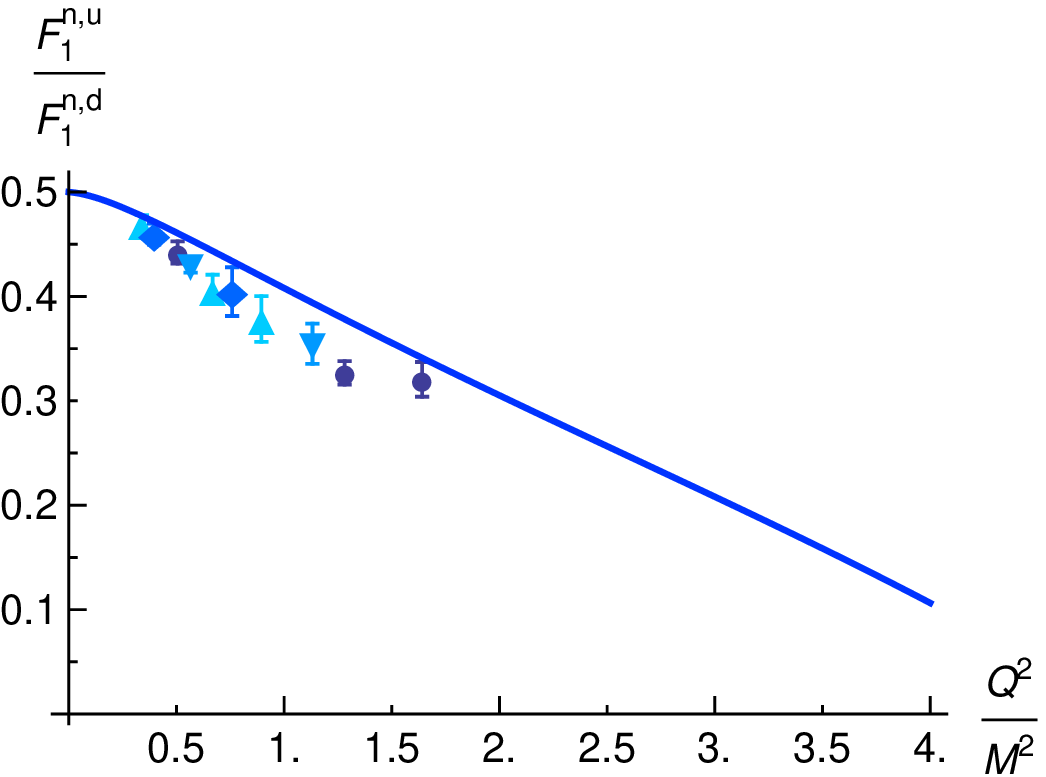}

\includegraphics[clip,width=0.36\textheight]{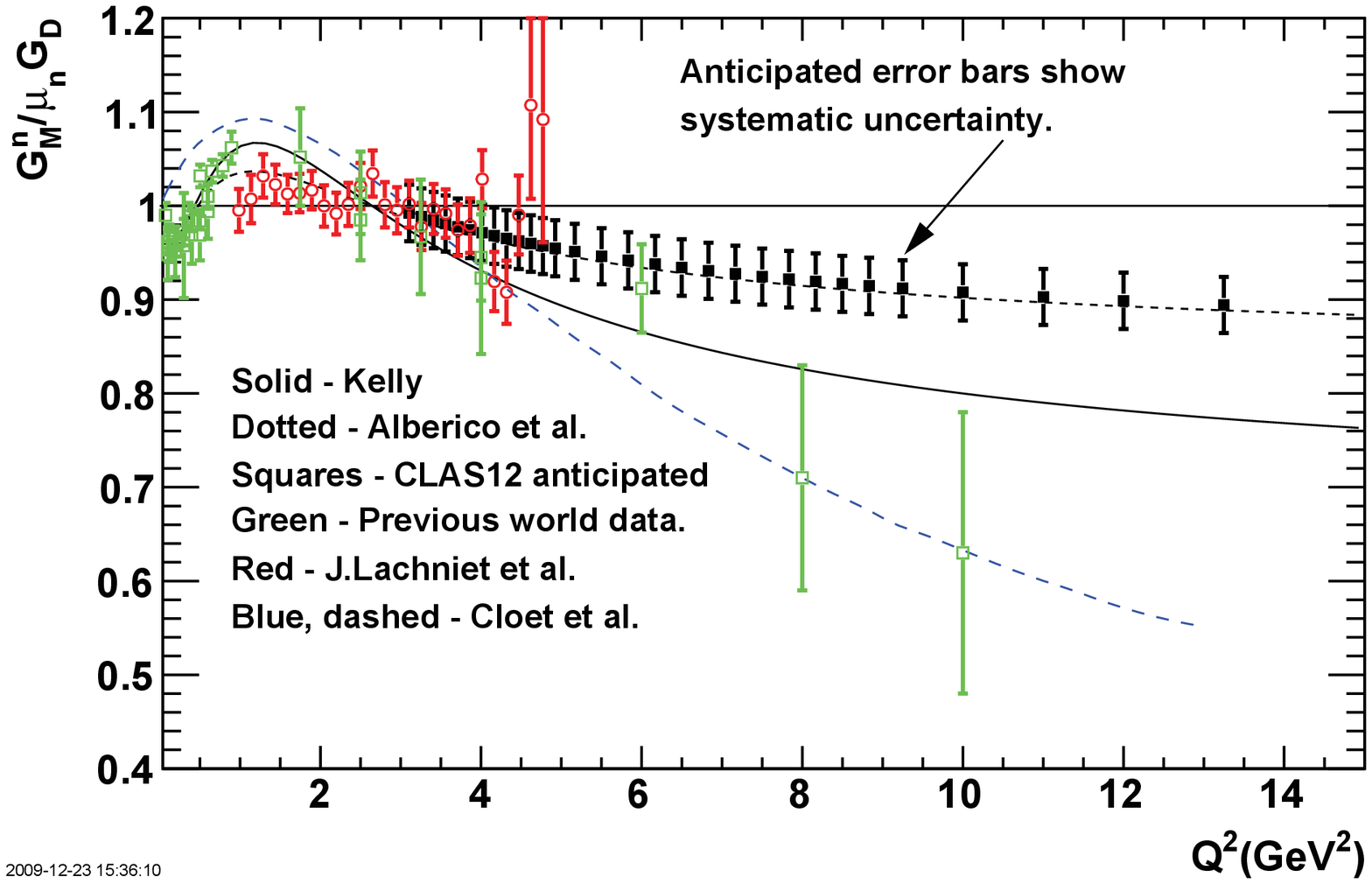}

\caption{\label{GMnGilfoyle}
\underline{Left panel}.  Computed ratio of flavour-separated contributions to neutron's Dirac form factor ($M =\,$nucleon mass) \emph{solid curve}.  Data, reconstructed from selected neutron electric form factor data using parametrisations of Ref.\,\protect\cite{Kelly:2004hm} when necessary \protect\cite{SRiordan}:
Ref.\,\protect\cite{Glazier:2004ny}, \emph{up-triangles};
Ref.\,\protect\cite{Bermuth:2003qh}, \emph{diamonds};
Ref.\,\protect\cite{Plaster:2005cx}, \emph{circles};
and Ref.\,\protect\cite{Warren:2003ma}, \emph{down-triangles}.
\underline{Right panel}.  Neutron magnetic form factor divided by dipole fit: \emph{long-dashed curve}, DSE prediction \protect\cite{Cloet:2008re}; \emph{solid curve}, \protect\cite{Kelly:2004hm}; \emph{open circles}, recent CLAS results \protect\cite{Lachniet:2008qf} (red) and pre-2008 data (green);
and black \emph{squares}, anticipated $Q^2$-coverage and errors for an experiment at the upgraded JLab \cite{JGilfoyle} -- the magnitude of these points is naturally arbitrary at this time but they were placed to follow a recent fit to now-extant experimental data (\emph{short-dashed curve}, \protect\cite{Alberico:2008sz}).
}
\end{figure}

In the left panel of Fig.\,\ref{GMnGilfoyle} we depict a ratio of flavour-separated contributions to the neutron's Dirac form factor.  The predicted $Q^2$-dependence owes to the presence of axial-vector diquark correlations in the nucleon.  It has been found \cite{Cloet:2008re} that the neutron's singly-represented $u$-quark is more likely to be struck in association with an axial-vector diquark correlation than with a scalar, and form factor contributions involving an axial-vector diquark are soft.  On the other hand, the doubly-represented $d$-quark is predominantly linked with harder scalar-diquark contributions.
NB.\ When isospin is a good symmetry, $F_1^{n,u}/F_1^{n,d} = F_1^{p,d}/F_1^{p,u}$.

In the right panel of Fig.\,\ref{GMnGilfoyle} we depict $G_M^n$ computed using the Faddeev equation \cite{Cloet:2008re}.  Given that it is common practice to compare nucleon form factors with an empirical dipole, the DSE result is presented in a similar way.  Namely, the function $1/(1+Q^2/m_D^2)^2$ was fitted to the DSE result on $2\leq Q^2/M_N^2 < 14$.  This domain excludes the region whereupon pion cloud effects are significant and maximises coverage of the domain on which the quark-core calculation is most reliable.  The fit produced $m_D= 1.05\, M_N$ cf.\ the empirical value, $m_D^{\rm emp}=0.90\,M_N$, which is just 14\% smaller.

The DSE predictions in Fig.\,\ref{GMnGilfoyle} depend sensitively on the momentum-dependence of the dressed-quark mass function, which feeds into properties such as the diquarks' mass and size.  Since in the computation of elastic and transition form factors, the probe's input momentum $Q$ is principally shared equally amongst the dressed-quarks, then each absorbs a momentum fraction $Q/3$.  Thus, to scan the behaviour of the mass function on the domain $p^2\in [0.2,1.2]\,$GeV$^2$, upon which its character changes from nonperturbative to perturbative, one requires $Q^2\in [2,15]\,$GeV$^2$.  The figure illustrates that comparison and feedback between DSE results and forthcoming precision data on nucleon form factors can serve as a means by which to empirically chart the momentum evolution of the dressed-quark mass function, and therefrom the infrared behavior of QCD's $\beta$-function.  In particular, it should enable the unambiguous location of the transition boundary between the constituent- and current-quark domains that is signalled by the sharp drop apparent in Fig.\,\ref{gluoncloud} and which can likely be related to an infrared inflexion point in QCD's running coupling, whose properties are determined by the $\beta$-function.

\section{Epilogue}
Dynamical chiral symmetry breaking (DCSB) is a fact in QCD.  It is manifest in dressed-propagators and vertices, and, amongst other things, it is responsible for:
the transformation of the light current-quarks in QCD's Lagrangian into heavy constituent-like quark's, in terms of which order was first brought to the hadron spectrum;
the unnaturally small values of the masses of light-quark pseudoscalar mesons;
the unnaturally strong coupling of pseudoscalar mesons to light-quarks -- $g_{\pi \bar q q} \approx 4.3$;
and the unnaturally strong coupling of pseudoscalar mesons to the lightest baryons -- $g_{\pi \bar N N} \approx 12.8 \approx 3 g_{\pi \bar q q}$.

Herein we have illustrated the dramatic impact that DCSB has upon observables: the spectrum, Secs.\,\ref{spectrum1} and \ref{spectrum2}; hadron form factors, Secs.\,\ref{FF1} and \ref{FF3}; and parton distribution functions, Sec.\,\ref{FF2}.  A ``smoking gun'' for DCSB is the behaviour of the dressed-quark mass function.  The momentum dependence manifest in Fig.\,\ref{gluoncloud} is an essentially quantum field theoretical effect.  Exposing and elucidating its effects therefore requires a nonperturbative and symmetry-preserving approach, where the latter means preserving Poincar\'e covariance, chiral and electromagnetic current-conservation, etc.  The Dyson-Schwinger equations (DSEs) provide such a framework.  Experimental and theoretical studies are underway that will identify observable signals of $M(p^2)$ and thereby explain the most important mass-generating mechanism for visible matter in the Standard Model.

There are many reasons why this is an exciting time in hadron physics.  We have focused on one.  Namely, through the DSEs, we are positioned to unify phenomena as apparently diverse as the: hadron spectrum; hadron elastic and transition form factors, from small- to large-$Q^2$; and parton distribution functions.  The key is an understanding of both the fundamental origin of nuclear mass and the far-reaching consequences of the mechanism responsible; namely, DCSB.  These things might lead us to an explanation of confinement, the phenomenon that makes nonperturbative QCD the most interesting piece of the Standard Model.


\begin{theacknowledgments}
We acknowledge valuable input from A.~Bashir, I.\,C.~Clo\"et, J.~Gilfoyle, L.~X.~Guti\'errez-Guerrero, A.~K{\i}z{\i}lers\"u, Y.-X.~Liu, S.~Riordan and P.~C.~Tandy.
CDR thanks the participating staff and students in the Department of Physics, University of Sinaloa, Culiac\'an, for their assistance and hospitality during the Workshop
and the preceding Mini-Courses.
This work was supported by:
the National Natural Science Foundation of China, contract no.\ 10705002;
the U.\,S.\ Department of Energy, Office of Nuclear Physics, contract no.~DE-AC02-06CH11357;
and the U.\,S.\ National Science Foundation, under grant no.\ PHY-0903991, in conjunction with a CONACyT Mexico-USA collaboration grant.
\end{theacknowledgments}



\bibliographystyle{aipproc}   

\end{document}